%% file: DMFridge_v15.tex
\documentclass[
    twocolumn,
	prd,
	amssymb,
	preprintnumbers,
	secnumarabic,
	nofootinbib]{revtex4-1}

\pdfoutput=1

\usepackage{graphicx}
\usepackage{enumitem}
\usepackage{latexsym}
\usepackage{amsfonts}
\usepackage{amssymb}
\usepackage{color}
\usepackage{units}
\usepackage{amsmath}
\usepackage{slashed}
\usepackage{dcolumn}
\usepackage{verbatim}
\usepackage{float}
\usepackage{multirow}
\usepackage{xspace}
\usepackage[normalem]{ulem}
\usepackage{url}

\renewcommand{\i}{\text{i}}
\newcommand{\dd}{\text{d}}

\newcommand{\mdm}{m_{\chi}}

\usepackage[pdfauthor=Nirmal Raj]{hyperref}

\input{universalnewcommands.tex}

\def\mmed{m_\phi}

\newcommand{\rg}[1]{\textcolor{blue}}

\allowdisplaybreaks

\begin{document}
\preprint{ULB-TH/20-12}
\title{Observing the thermalization of dark matter in neutron stars}

\author{Raghuveer Garani} \email{rgaranir@ulb.ac.be}
\affiliation{Service de Physique Theorique, Universite Libre de Bruxelles,
Boulevard du Triomphe, CP225, 1050 Brussels, Belgium}

\author{Aritra Gupta} \email{aritra.gupta@ulb.be}
\affiliation{Service de Physique Theorique, Universite Libre de Bruxelles,
Boulevard du Triomphe, CP225, 1050 Brussels, Belgium}

\author{Nirmal Raj} \email{nraj@triumf.ca}
\affiliation{TRIUMF, 4004 Wesbrook Mall, Vancouver, BC V6T 2A3, Canada}

\date{\today}

\begin{abstract}
A promising probe to unmask particle dark matter is to observe its effect on neutron stars, the prospects of which depend critically on whether captured dark matter thermalizes in a timely manner with the stellar core via repeated scattering with the Fermi-degenerate medium.
In this work we estimate the timescales for thermalization for multiple scenarios.
These include: (a) spin-0 and spin-$\frac{1}{2}$ dark matter, 
(b) scattering on non-relativistic neutron and relativistic electron targets accounting for the respective kinematics,
(c) interactions via a range of Lorentz-invariant structures,
(d) mediators both heavy and light in comparison to the typical transfer momenta in the problem.
We discuss the analytic behavior of the thermalization time as a function of the dark matter and mediator masses, and the stellar temperature.
Finally, we identify parametric ranges where both stellar capture is efficient and thermalization occurs within the age of the universe. 
For dark matter that can annihilate in the core, these regions indicate parametric ranges that can be probed by upcoming infrared telescopes observing cold neutron stars.
\end{abstract}

\maketitle

\section{Introduction}

By virtue of their steep gravitational potentials accelerating ambient particles to semi-relativistic speeds, and their large densities enabling efficient capture, compact stars have served as valuable laboratories for studying dark matter (DM)~\cite{Goldman:1989nd,Gould:1989gw,
  Kouvaris:2007ay, 
  Bertone:2007ae, 
  McCullough:2010ai, 
  deLavallaz:2010wp, 
  Kouvaris:2010jy,
  Kouvaris:2010vv, 
  Kouvaris:2011fi, 
  McDermott:2011jp, 
  Guver:2012ba, 
  Capela:2013yf,
  Bertoni:2013bsa,
  Bramante:2013hn, 
  Bramante:2013nma, 
  Bell:2013xk, 
  Perez-Garcia:2014dra,
  Graham:2015apa,
  Bramante:2015cua,
  Cermeno:2016olb, 
  Krall:2017xij,
  Ellis:2017jgp, 
  Ellis:2018bkr,
  Kouvaris:2018wnh,
  Graham:2018efk, 
  McKeen:2018xwc,
  Acevedo:2019gre,
  Janish:2019nkk, 
  Dasgupta:2019juq,
  Lin:2020zmm,
  Dasgupta:2020dik,
  Genolini:2020ejw,
  Horowitz:2020axx,
  Dasgupta:2020mqg}. 
  Neutron stars (NS) in particular could provide wide-ranging constraints through one of the following means. 
  {\bf (a)} A heat signature in old, isolated, nearby NSs that may be observed by imminent infrared telescopes~\cite{Baryakhtar:2017dbj}. 
  The NS luminosity is sourced by the kinetic energy of infalling DM~\cite{Baryakhtar:2017dbj,Raj:2017wrv,
  Garani:2018kkd,
   Acevedo:2019agu,
  Bell:2018pkk,
  Chen:2018ohx, 
  Camargo:2019wou,
  Hamaguchi:2019oev,
  Bell:2019pyc,
  Garani:2019fpa,
  Joglekar:2019vzy,
  Joglekar:2020liw,
  Bell:2020jou}, and in some scenarios, additionally by the annihilation of captured DM as well.
  {\bf (b)} Their very existence, since in certain models involving non-annihilating ``asymmetric" DM, gravitational collapse may be triggered in the core and a star-destroying black hole may be formed.
  For captured DM to either collect and annihilate efficiently in the stellar core or to form a black hole, it must first thermalize with the core via repeated scattering within timescales of interest.
   This is a model-dependent process, and has direct bearing on observational prospects.
 
 NS-DM thermalization was studied in detail in Ref.~\cite{Bertoni:2013bsa}, but in a setting restricted to spin-$\half$ DM and vector-vector contact interactions with Standard Model (SM) fermions.
 With the discovery and observation of relevant NSs imminent~\cite{Baryakhtar:2017dbj,Raj:2017wrv}, and no clarity yet on the identity of DM, it is timely to expand the thermalization program to a broader range of DM scenarios.
  In this study, we compute the thermalization timescales, accounting for Pauli-blocked phase space, for spin-0 and spin-$\half$ DM interacting via various Lorentz-invariant structures with neutron and electron targets in the stellar core, as well as treat the case of ``light mediators" with masses smaller than the momentum transfers involved in thermalization.  
  We then discuss the interplay of these thermalization timescales with NS capture sensitivities.
  This study thereby identifies target regions for infrared telescope astronomers looking for DM annihilation heating of NSs.
  Several such regions are identified, hence this work furthers the case for constraining the thermal luminosity of candidate NSs. 
  
  This paper is organized as follows.
  In Section~\ref{sec:therm} we review some general considerations in computing energy loss rates and thermalization in the stellar core for both non-relativistic neutron and relativistic electron targets.
  We also outline the DM-SM interaction structures we consider.
  In Section~\ref{sec:results} we provide the thermalization timescales for all our scenarios, and discuss their analytic behavior with respect to DM mass, mediator mass, and the core temperature. 
  We also identify regions in the space of DM and mediator masses that are promising for upcoming astronomical observations.
  In Section~\ref{sec:concs} we summarize our findings and discuss the future scope of our work.
  In the appendices we collect technical details of our estimation of the thermalization timescales.
  
%%%%%%%%%%%
\section{Thermalization: general considerations}
\label{sec:therm}
%%%%%%%%%%%

In treating DM-NS thermalization we implicitly assume that incident DM is gravitationally captured by the NS.
In Sec.~\ref{sec:results} we will check this assumption explcitly and discuss the phenomenological implications.

Captured DM particles thermalize with the NS by losing energy via repeated scattering with particles in the medium.
The time taken to achieve thermalization may be divided into two epochs. 
In the first epoch of interval $t_1$, the orbits of gravitationally bound DM particles, which are highly eccentric, shrink as the DM loses energy during transits, until they become confined to within the NS radius. In the second epoch of interval $t_2$, the confined DM particles scatter further with the NS medium; their orbits continue to shrink down to the ``thermal radius" determined by the NS core density and temperature. 
When this is completed the DM kinetic energy equals the NS temperature $T_{\rm NS}$. 
As shown in Refs.~\cite{Kouvaris:2010jy,Garani:2018kkd}, $t_2$ exceeds $t_1$ by a few orders of magnitude, thus the total thermalization timescale is determined by the second epoch.  
Consequently, in this work we focus solely on this second epoch. 

The interaction rate of DM with fermion degenerate matter is described by Fermi's golden rule that accounts for the phase space available for scattering. 
Consider the elastic scattering of a DM particle $\chi$ off a distribution of target particles $T$: 
$\chi (k) + T (p) \rightarrow \chi (k^\prime) + T (p^\prime)$.
The interaction rate per DM particle reads~\cite{Bertoni:2013bsa}:
%%%%
\bea
\label{eq:rate}
&&d \Gamma =2 \frac{\dd^3 k^\prime}{(2 \pi)^3} S(q_0,q)~, \nonumber\\
\nn &&S(q_0,q)=\int \frac{\dd^3 p^\prime}{(2 \pi)^3 2 E_{p^\prime} 2 E_{k^\prime}}  \int \frac{\dd^3 p}{(2 \pi)^3 2 E_{p} 2 E_k}  \times  \\
&&(2 \pi)^4 \delta^4\left(k+p -k^\prime -p^\prime \right) |\mathcal{M}|^2 f(E_p) \left(1 -f(E_{p^\prime})  \right)~,
\eea
%%%
where the second line above is the response function.
The Fermi distribution function of the target particles with chemical potential $\mu$ is:
%%%
\beq
f(E_p) = \frac{1}{\exp((E_p-\mu)/T_{\rm NS})+ 1}.
\eeq
%%%

The rate of energy loss is given by
%%%
\beq
\Phi = \int \dd \Gamma \times (E_i - E_f)~,
\label{eq:Elossrate}
\eeq
%%%
where $k'$ is integrated from 0 to $k$.
Using this we can write down the time taken to thermalize with the NS, i.e. to reach a final energy $E_f = 3/2\, \,T_{\rm NS}$ starting with an initial energy $E_0 =  m_\chi v_{\rm esc}^2/2$:
%%%%
\bea
\label{eq:ttime}
\tau_{\rm therm} = - \int^{E_f}_{E_0} \frac{\dd E_i}{\Phi}.
\eea
%%%%

We thus see that the key quantity determining thermalization is the response function, which in turn depends on $\chi$-$T$ interaction structure via the squared amplitude $|\mathcal{M}|^2$.
We now inspect these ingredients in more detail.
\vspace{.5in}

%%%%%%%%%
\subsection{The response function}
%%%%%%%%%

Integrating Eq.~\eqref{eq:rate} over $p'$, the response function becomes
\bea
S(q_0,q)&=&\int \frac{\dd^3 p}{(2 \pi)^2 2 } \frac{|\mathcal{M}|^2} { 16 E_{p^\prime}  E_{k^\prime} E_{p}  E_k}  \times \nonumber \\
&&   \delta\left(q_0 - E_p + E_{p^\prime}\right)  f(E_p) \left(1 -f(E_{p^\prime})  \right), 
\label{eq:Sred}
\eea
where the $\delta^3$ has simply enforced momentum conservation, $\mathbf{q} = \mathbf{p^\prime} -\mathbf{p}$.
While DM particles move at semi-relativistic speeds during NS capture, thermalization occurs after they have slowed down to non-relativistic speeds, thus we always set $E_k = E_{k^\prime} = m_\chi$.
On the other hand, the target could be either relativistic or non-relativistic.
Neutrons (as well as protons and muons) in the stellar core typically have Fermi momenta $p_{\rm F}$ smaller than their rest mass $m_{T}$ and are thus expected to be non-relativistic, whereas electrons are highly degenerate ($p_{\rm F} \gg m_T$) and hence ultra-relativistic.
We consider DM scattering on both neutrons and electrons in the core.

For non-relativistic scattering targets, we have $E_p = m_T + p^2/(2 \,m_T)$. 
Assuming the squared amplitude depends only on the Mandelstam variable $t$ \footnote{If the squared amplitude depends on both $t$ and $s$, integration over the azimuthal angle in Eq.~\eqref{eq:Sred} is non-trivial and analytic results are not easily obtained. 
Since $t$-only-dependent squared amplitudes already span the range of interesting behavior for thermalization, we study only interactions giving rise to them.}, the response function in the limit $\mu \gg T_{\rm NS}$ is obtained, following the treatment in Ref.~\cite{Reddy:1997yr}, as:
%%%%
\bea
\label{eq:response-nr-ex}
S^{\rm non-rel}(q_0,q)&=& \frac{|\mathcal{M}|^2} { 16 \pi m^2_\chi } \frac{q_0}{q}\Theta\left(\mu - \frac{1}{4} \frac{(q_0 -q^2/2 m_T)^2}{q^2/2 m_T}\right)~, \nonumber \\
\eea
%%%%
where $q \equiv |\mathbf{q}|$, a notation we use for the remainder of the paper. 
Note that we recover the approximate response function quoted in Refs.~\cite{Reddy:1997yr,Bertoni:2013bsa}
by rewriting the Heaviside theta function above in terms of $q_0$, and by keeping only the leading order term in the Fermi velocity ($v_{\rm F} = \sqrt{2\, \mu/m_T}$):
%%%%
\bea
\label{eq:response-nr1}
S^{\rm non-rel}_{\rm approx}(q_0,q)&\approx& \frac{|\mathcal{M}|^2} { 16 \pi m^2_\chi } \frac{q_0}{q}\Theta(q v_f - q_0).
\eea
%%%%
 This form of the response function holds true only in the deeply non-relativistic limit, whereas Eq.~\eqref{eq:response-nr-ex} is valid also in the quasi-relativistic regime, thus allowing us to smoothly transition from the non-relativistic to the relativistic limit. 
 Despite these analytical differences in the response functions, as we shall see later, for DM-NS thermalization they matter little.  

 Next we examine relativistic electron targets.
 It is expected that electrons inside old NSs are completely relativistic~\cite{Cohen1970,Potekhin:2013qqa,Goriely:2013xba}, with typical chemical potentials $\mu_F = \mathcal{O}(0.1)$ GeV. 
 In this limit, integrating Eq.~\eqref{eq:rate} over $p',p$ gives the following response function in the $\mu \gg T_{\rm NS}$ limit, as derived in Appendix~\eqref{app:rel}.
\bea
\label{eq:response-rel}
S^{\rm rel}(q_0,q)&=& \frac{|\mathcal{M}|^2}{16\,\pi\,m^2_\chi} \frac{q_0}{q} \Theta(2 \mu +   q_0 - q ). 
\eea

%%%%%%%
\begin{figure}
  \includegraphics[width=0.49\textwidth]{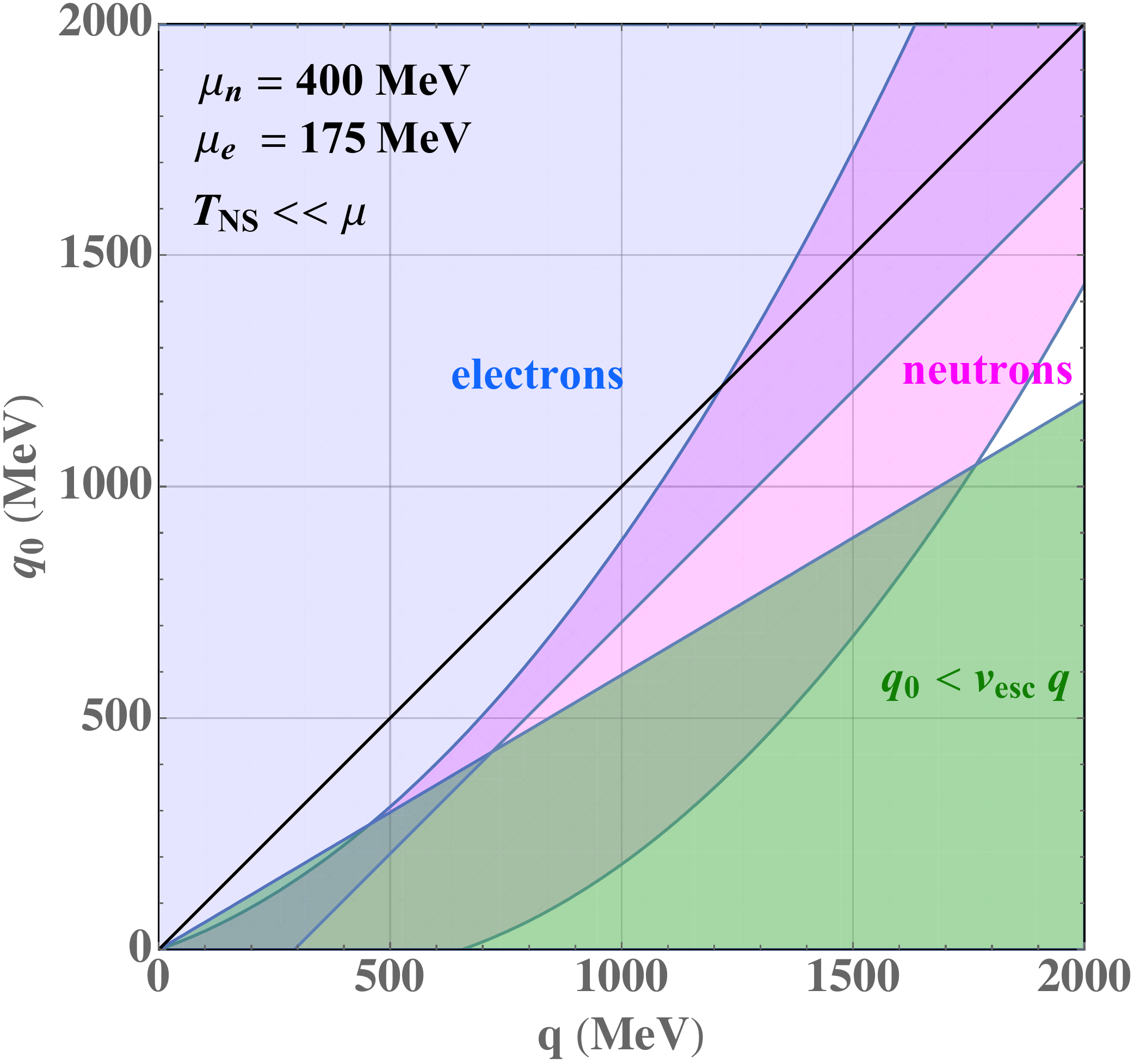}            
	\caption{Regions of phase space allowed in the response functions in Eqs.~\eqref{eq:response-nr-ex} and \eqref{eq:response-rel} for neutron and electron targets, encapsulating non-relativistic and relativistic kinematics respectively, as well as Pauli-blocking.
 Also shown is the region $q_0 < v_{\rm esc} q$, depicting dark matter kinematics after capture. 
During dark matter thermalization with the neutron star core, scatters only occur where the above two regions overlap.
The timescale for the thermalization process is set by the last few scatters with $q, q_0$ much smaller than target masses, i.e. by the bottom left region of this plot.}
	\label{fig:response}
\end{figure}
%%%%%%%

We notice that the response function for neutron and electron targets have similar forms, except for the theta functions that enforce Pauli-blocking in non-relativistic and relativistic regimes, respectively .
To understand this phase space restriction better, we display in Fig.~\ref{fig:response} regions where the response functions in Eqs.~\eqref{eq:response-nr-ex} and \eqref{eq:response-rel} are non-zero. 
We also show the region $q_0 < v_{\rm esc} q$, which depicts the phase space occupied by DM after capture.
Here $v_{\rm esc} = 0.6 c$ is the surface escape speed of a typical NS. 
For reference, the diagonal black line denotes $q_0 = q$. 
Successful scattering can occur in regions that overlap with DM phase space. 

The allowed regions in the response functions could be further understood through the following.
During thermalization, DM particles (targets) continuously lose (gain) energy, i.e. $q_0 \ge 0$. 
Thus from Eqs.~\eqref{eq:response-nr-ex} and \eqref{eq:response-rel} we have the condition $q < \sqrt{8\,\mu\,m_T}$ for successful scattering on neutrons, and similarly $q< 2\mu$ for electrons.
These conditions imply that scattering in all directions is allowed for neutron targets, whereas scattering could be restricted to forward directions for electron targets.
To see this, consider the scattering angle in neutron scattering, $\cos\theta_{pq} = (2 m_T q_0 - q^2)/(2\,pq)$, obtained from energy conservation.
Using $q_0 \geq 0$ and the above condition, we have $\cos\theta_{pq} > -\sqrt{2\,\mu_n\,m_T}/p$.
For neutron targets the numerator is $\Oc(0.1)$ GeV, whereas $p$ is much smaller, hence this inequality is weaker than the condition $\cos\theta_{pq} > -1$. 
 On the other hand, for electron targets we have $\cos\theta_{pq} = (2\,pq)^{-1} (q^2_0 + 2pq_0 - q^2)$, implying $\cos\theta_{pq} > -\mu_e/p$.
 Since $p \sim \mu_e$, this condition could be more restrictive on backward scattering.

These considerations, however, have little effect on our final estimates.
The timescale for thermalization is set primarily by the last stages of the process, involving dark matter kinetic energies approaching the stellar temperature.
It can be seen, most clearly in the analytic expressions in Appendix~\ref{app:non-rel}, that the rate of energy loss in Eq.~\eqref{eq:Elossrate} is smallest for these last scatters, and therefore that the times taken for these are the longest. 
In this regime $q$, $q_0$ are much smaller than $\mu$ and target masses, i.e. the phase space here is the bottom left corner of Fig.~\ref{fig:response}, where all response functions overlap with the DM phase space. In this region the theta functions in Eqs.~\eqref{eq:response-nr-ex} and \eqref{eq:response-rel} are trivially unity.    

As discussed in Ref.~\cite{Bertoni:2013bsa}, the above treatment of DM kinematics and Fermi-degenerate phase space yields non-trivial results that are otherwise missed by simpler treatments of Pauli-blocking such as in Ref.~\cite{McDermott:2011jp}. 
For instance, for momentum-independent $|\mathcal{M}|^2$, $\tau_{\rm therm}$ scales as $\mdm^{-1}T^{-2}_{\rm NS}$ for $\mdm \gg m_T$ in our treatment (as derived below).
This differs from the result of Ref.~\cite{McDermott:2011jp}: $\tau_{\rm therm}\propto \mdm^{2}T^{-1}_{\rm NS}$. 
Not only does the thermalizaion time decrease as a different power of the NS temperature in our result, but counter-intuitively, it {\em decreases} as the DM mass (hence initial kinetic energy) is increased.
This behavior is only captured by the full treatment as above.
Another counter-intuitive result in our treatment is the total number of scatters that DM undergoes before thermalizing with the NS.
As estimated in Appendix~\ref{app:Nscatt}, it only takes $\Oc(10-100)$ scatters for the DM kinetic energy to fall from $ \mdm v^2_{\rm esc}/2$ to $T_{\rm NS} \simeq 0.1-10$ eV, whereas the calculation of Ref.~\cite{McDermott:2011jp} predicts that the number of scatters could be orders of magnitude more.

%%%%%%%
\subsection{Equation of state and neutron star properties}
\label{subsec:deos}
%%%%%%%

 The macroscopic properties of a NS such as its mass and radius, as well as thermoydnamic quantities like the chemical potential of constituent fermions (neutrons, protons, electrons and muons), are estimated from the equation of state (EoS) of matter at nuclear densities. 
 The EoS is derived from fits to available data at large finite densities~\cite{Potekhin:2013qqa,Goriely:2013xba}. 
 
 From the discussion above, the thermalization time is mostly independent of the chemical potential in the limit $\mu/T_{\rm NS} \gg 1$, as well as other macroscopic properties of the NS.
 As mentioned before, small energy and momentum transfers govern the thermalization time. 
 Physically, the target particles that participate in scattering are extremely close to their Fermi surface, with the minimum number of target particles participating in scattering proportional to $q_0\propto 2\, T_{\rm NS} >0$. 
 For a typical old NS with $T_{\rm NS} = 10^3-10^5~{\rm K} \simeq 10^{-10}-10^{-8}$~GeV and $\mu \sim 0.1$~GeV, we are always in the degenerate limit. 
 In other words, $\tau_{\rm therm}$ is a function of masses and $T_{\rm NS}$ only.

 For DM capture in the NS it is well known that the NS profile dependence of the capture rate could be as large as an order of magnitude~\cite{Baryakhtar:2017dbj,Garani:2018kkd,Bell:2020jou}.
 In presenting the interplay of our results with NS capture we choose the following benchmark predicted by the unified equation of state BSk-24~\cite{Pearson:2018tkr}:
 %%%%
 \bea
\nn M_{\rm NS} = 1.5~M_\odot,~~~R_{\rm NS} = 12.6~{\rm km}~ \\
\mu_n = 373~{\rm MeV},~~~\mu_e = 146~{\rm MeV}.
\label{eq:benchmarkNS}
 \eea
 %%%%
 The chemical potentials here are the volume-averaged values, which we adopt for our estimates of the capture rate in Sec.~\ref{sec:results}.

%%%%%%%
\begin{table} 
  \centering
  \begin{tabular}{lll}
  \hline
  	Name 
  	& Operator 
  	& $\sum_{\rm spins} |\mathcal{M}|^2$
  	\;
  	\\
  	 \hline
  	  \hline
   	$\mathcal{O}^\textnormal{F}_1$ 
  	& $ \left(\bar{\chi}\chi\right) \left(\bar{\xi}\xi\right)$ 
  	& $ \Lambda^{-4} \ \left(4m_\chi^2-t\right)\left(4m_T^2-t\right)$  
  	\\
  	$\mathcal{O}^\textnormal{F}_2$ 
  	& $\left(\bar{\chi} i\gamma^5\chi\right)\left(\bar{\xi}\xi\right)$ 
  	& $  \Lambda^{-4} \ t\left(t-4m_T^2\right)$
  	\\
  	$\mathcal{O}^\textnormal{F}_3$ 
  	& $%y_\xi 
  	\left(\bar{\chi}\chi\right)\left(\bar{\xi} i\gamma^5\xi\right)$ 
  	& $%y_\xi^2  
  	\Lambda^{-4} \
  	t\left(t-4m_\chi^2\right)$
  	\\
  	$\mathcal{O}^\textnormal{F}_4$ 
  	& $%y_\xi 
  	\left(\bar{\chi} i\gamma^5\chi\right)\left(\bar{\xi} i\gamma^5\xi\right)$  	
  	& $%y_\xi^2 
  	\Lambda^{-4} \
  	t^2$
  	\\
  	 \hline
  	%
  	%% BOSONS
  	%
   	$\mathcal{O}^\textnormal{S}_1$ 
  	& 
  	$%y_\xi 
  	\left(\chi^\dag\chi\right)
  	\left(\bar\xi\xi\right)$
  	& 
  	$
  	%y_\xi^2 
  	 \Lambda^{-2} \ 
  	\left(4m_{\rm T}^2-t\right)
  	$
  	\\
  	$\mathcal{O}^\textnormal{S}_2$ 
  	& 
  	$%y_\xi 
  	\left(\chi^\dag\chi\right)
  	\left( \bar\xi i\gamma^5 \xi \right)$
  	& $
  	%y_\xi^2
  	 \Lambda^{-2} \ \left(-t\right) 
  	$
  	  	\\
  	 \hline
  \end{tabular}   
  \caption{Effective contact operator interactions with cutoff $\Lambda$ between the dark matter field ($\chi$) and neutrons/electrons ($\xi$), and the corresponding squared tree-level scattering amplitudes.
  Operators with superscript F(S) correspond to spin-$\half$ (spin-0) DM.
  See Sec.~\ref{subsec:interxnstructs} for more details.
  }
  \label{tab:ampsq}
\end{table}
%%%%%%

%%%%%%%
\begin{figure*}
  \includegraphics[width=0.46\textwidth]{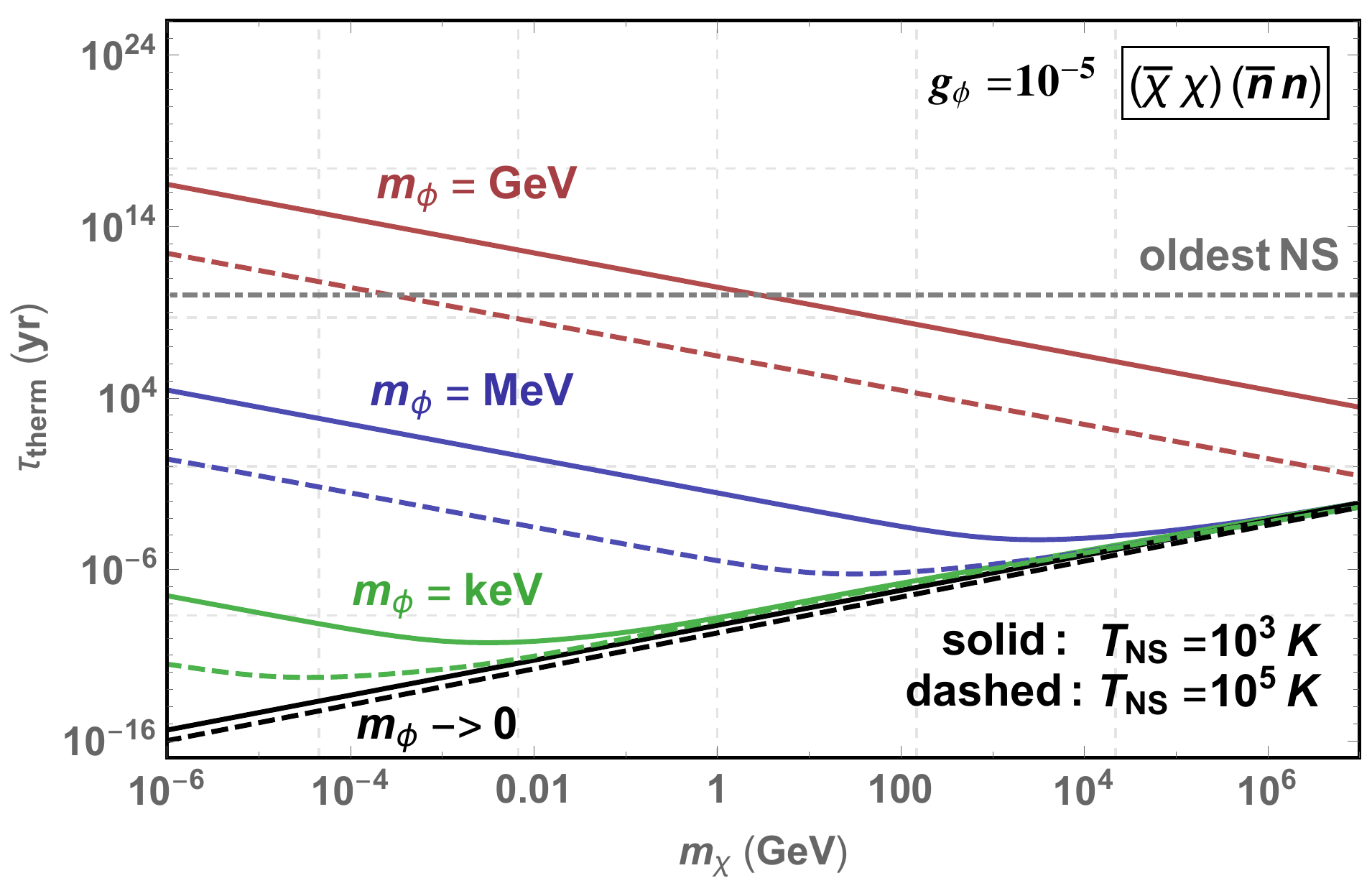}            
    \includegraphics[width=0.46\textwidth]{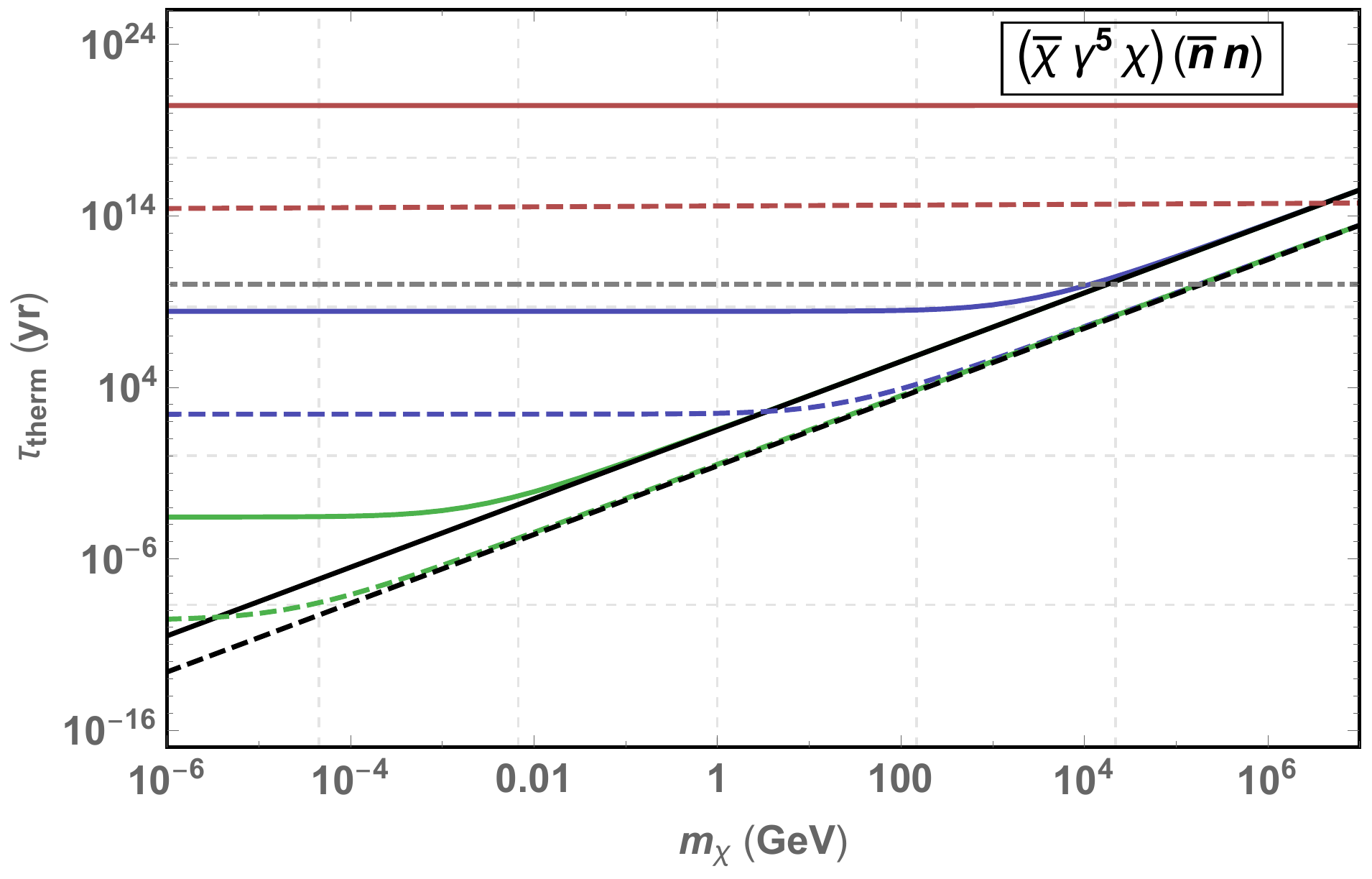} \\ 
    \includegraphics[width=0.46\textwidth]{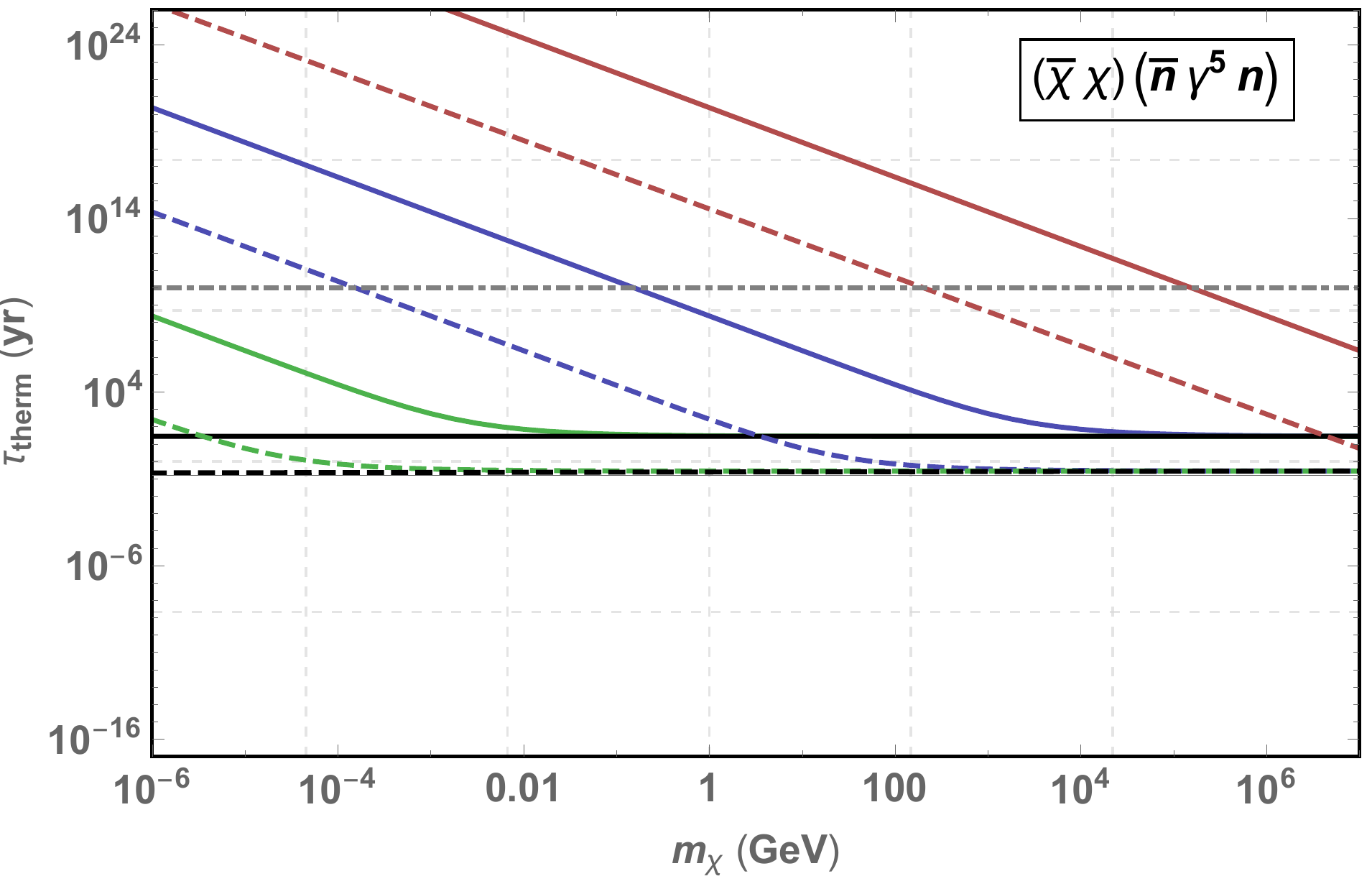}            
    \includegraphics[width=0.46\textwidth]{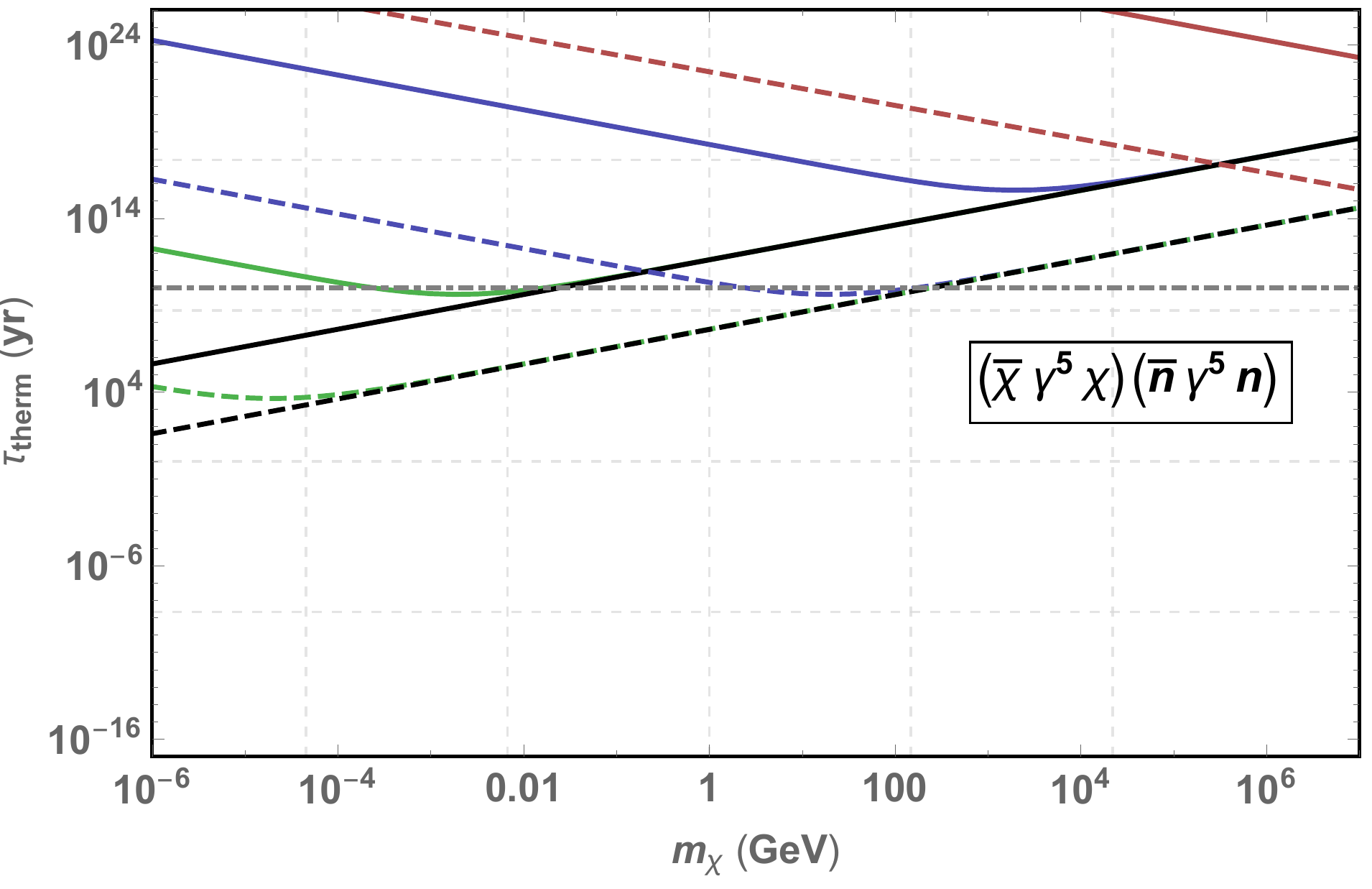} 
	\caption{Time taken for {\bf spin-$\half$ dark matter} to thermalize with neutron stars with temperatures $10^3$~K and $10^5$~K via repeated scattering on {\bf neutron} targets.
    Shown are the $\tau_{\rm therm}$ for mediator masses zero, keV, MeV, and GeV.  
    The panels correspond to interaction structures described in Sec.~\ref{subsec:interxnstructs} and Table~\ref{tab:ampsq}.
    The color code for all the panels is given by the first one.
    See text for further details.
	}
	\label{fig:tauthermnucleons}
\end{figure*}
%%%%%%%

%%%%%%%
\begin{figure*}
  \includegraphics[width=0.46\textwidth]{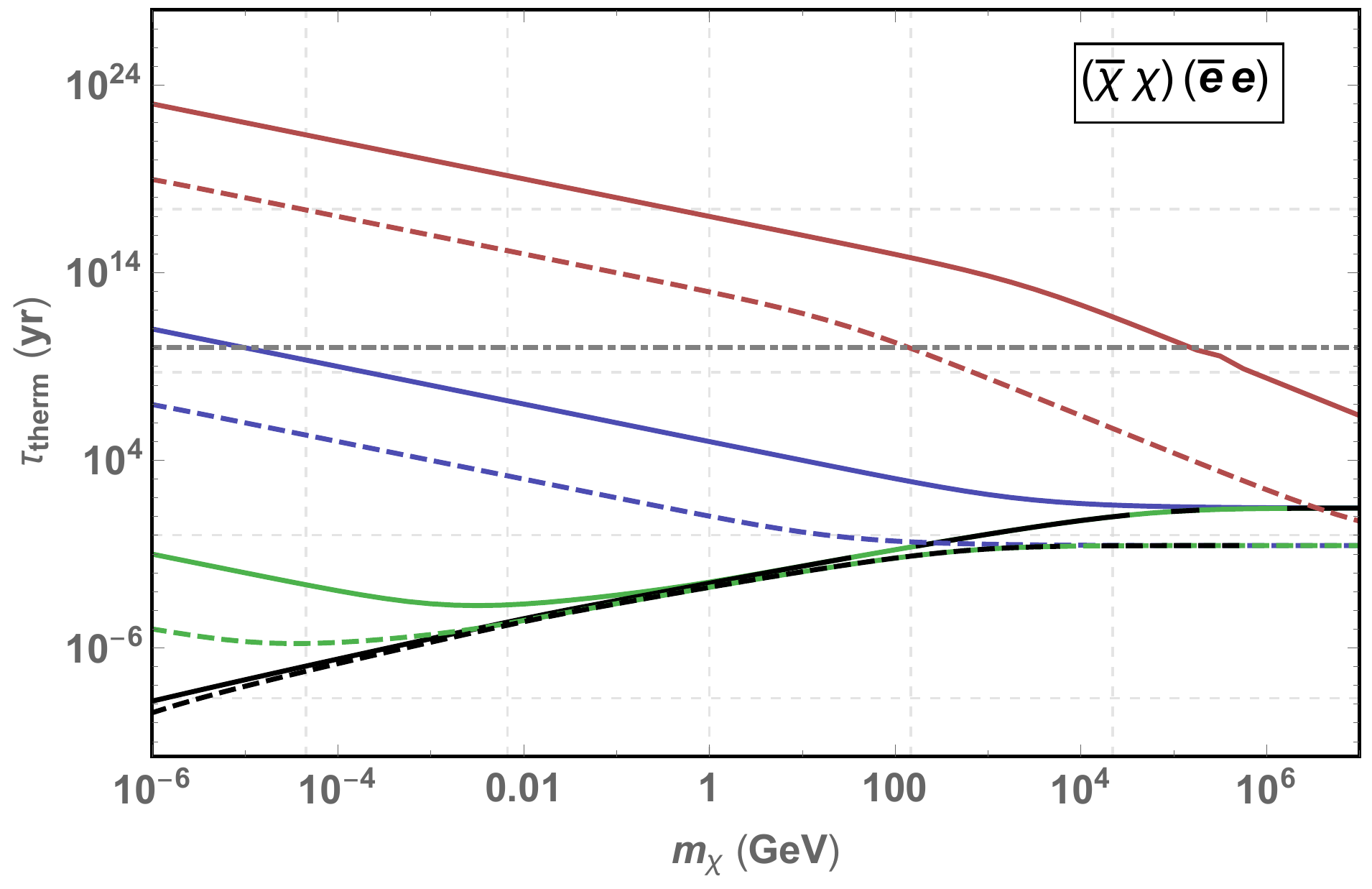}            
    \includegraphics[width=0.46\textwidth]{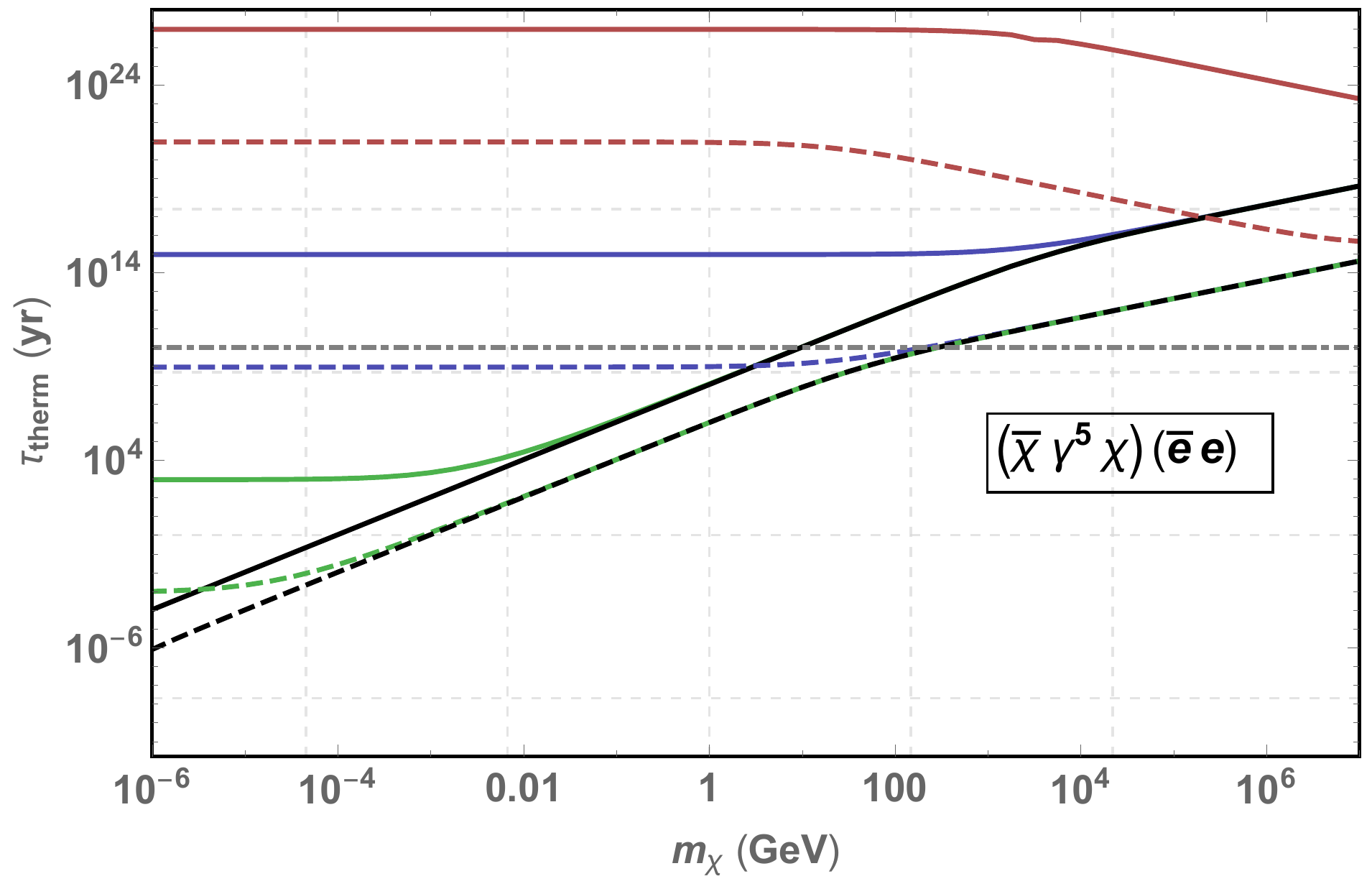} \\ 
	\caption{Same as Figure~\ref{fig:tauthermnucleons}, but for {\bf spin-$\half$ DM} and {\bf electron} targets.
   For the operators $\mathcal{O}^\textnormal{F}_3$ and $\mathcal{O}^\textnormal{F}_4$, $\tau_{\rm therm}$ is the same as in Figure~\ref{fig:tauthermnucleons}, and we omit displaying them.}
	\label{fig:tauthermelectrons}
\end{figure*}
%%%%%%%

%%%%%%%
\begin{figure*}
  \includegraphics[width=0.46\textwidth]{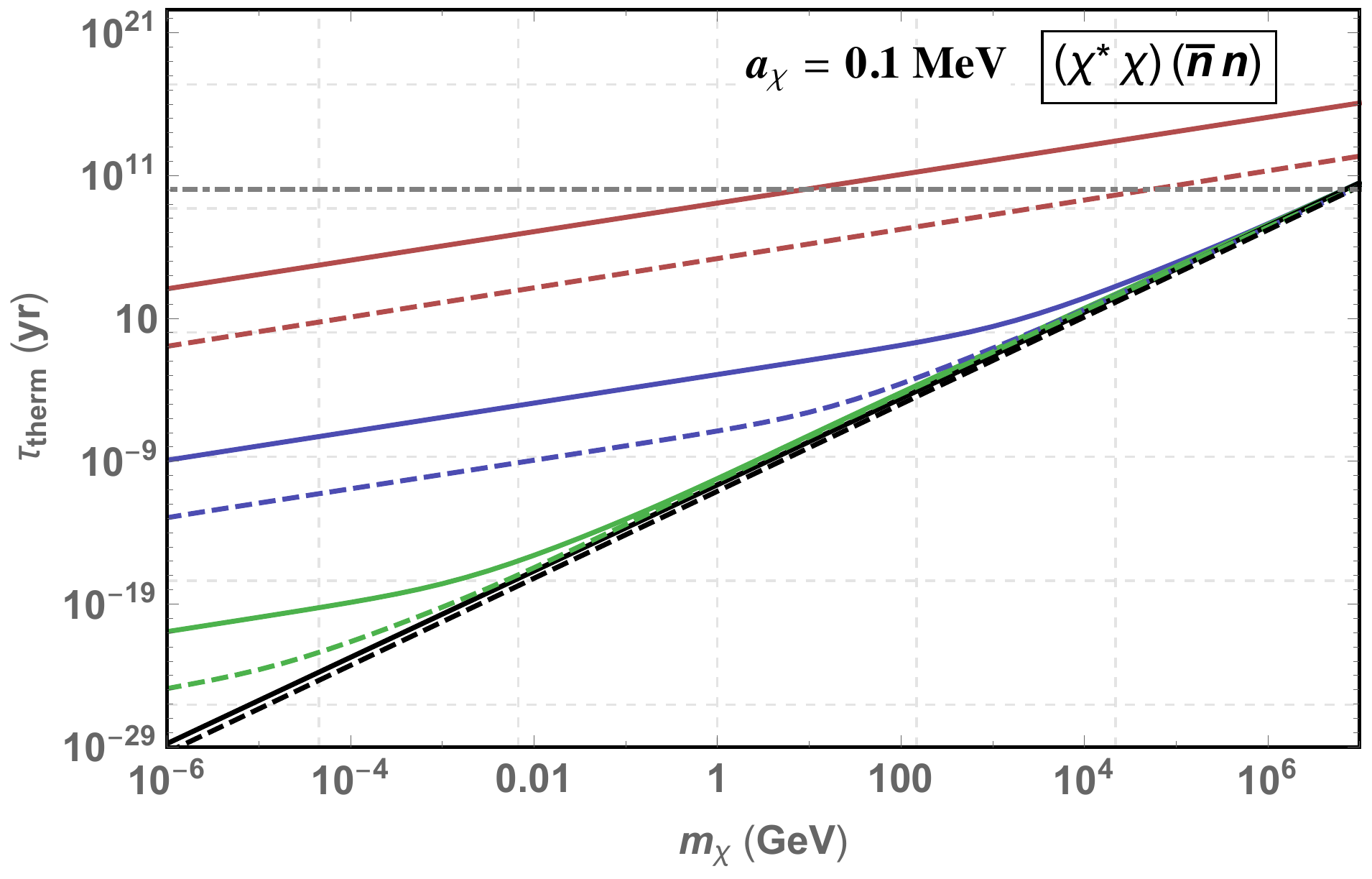}           
    \includegraphics[width=0.46\textwidth]{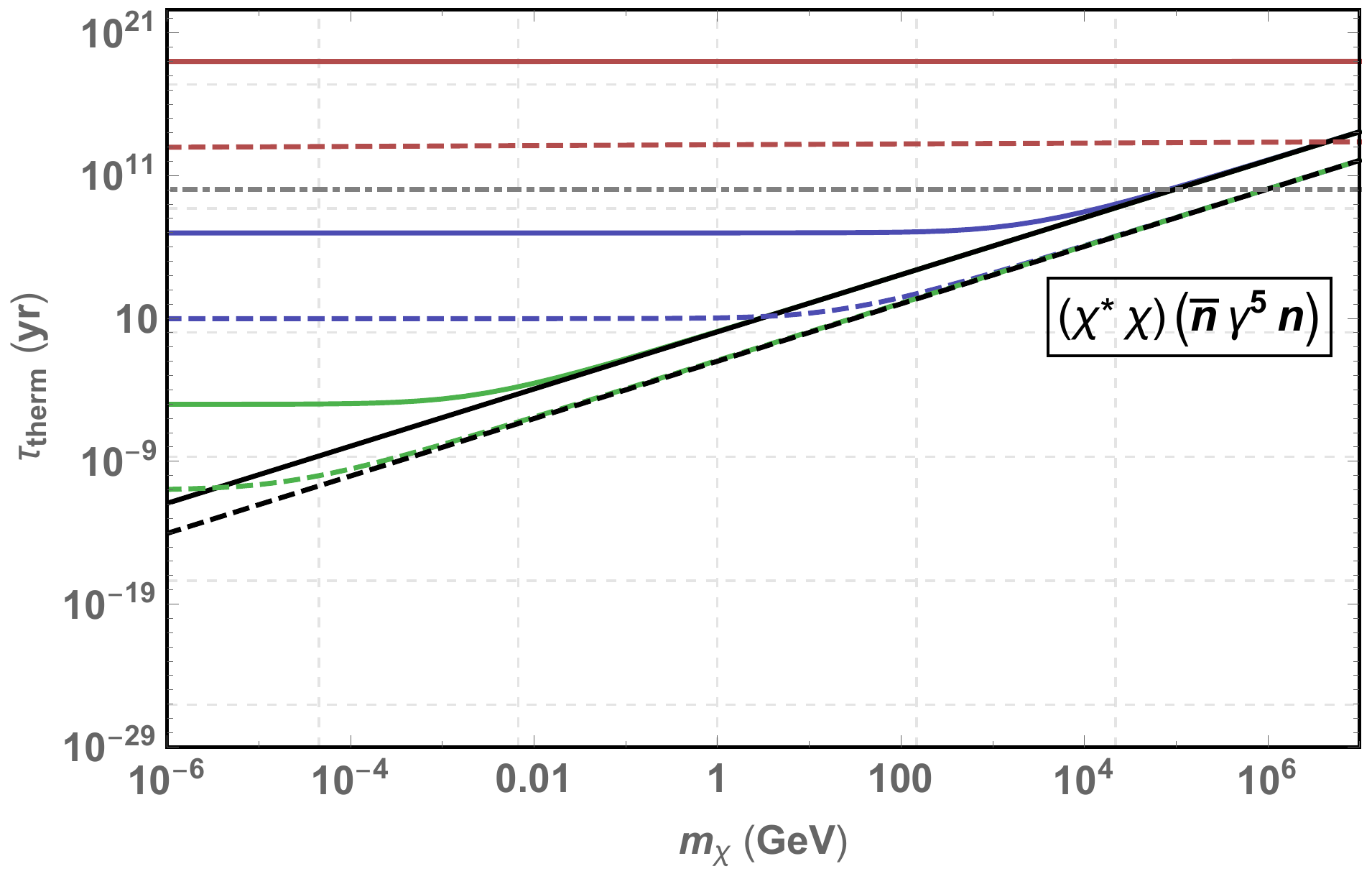} \\
    \includegraphics[width=0.46\textwidth]{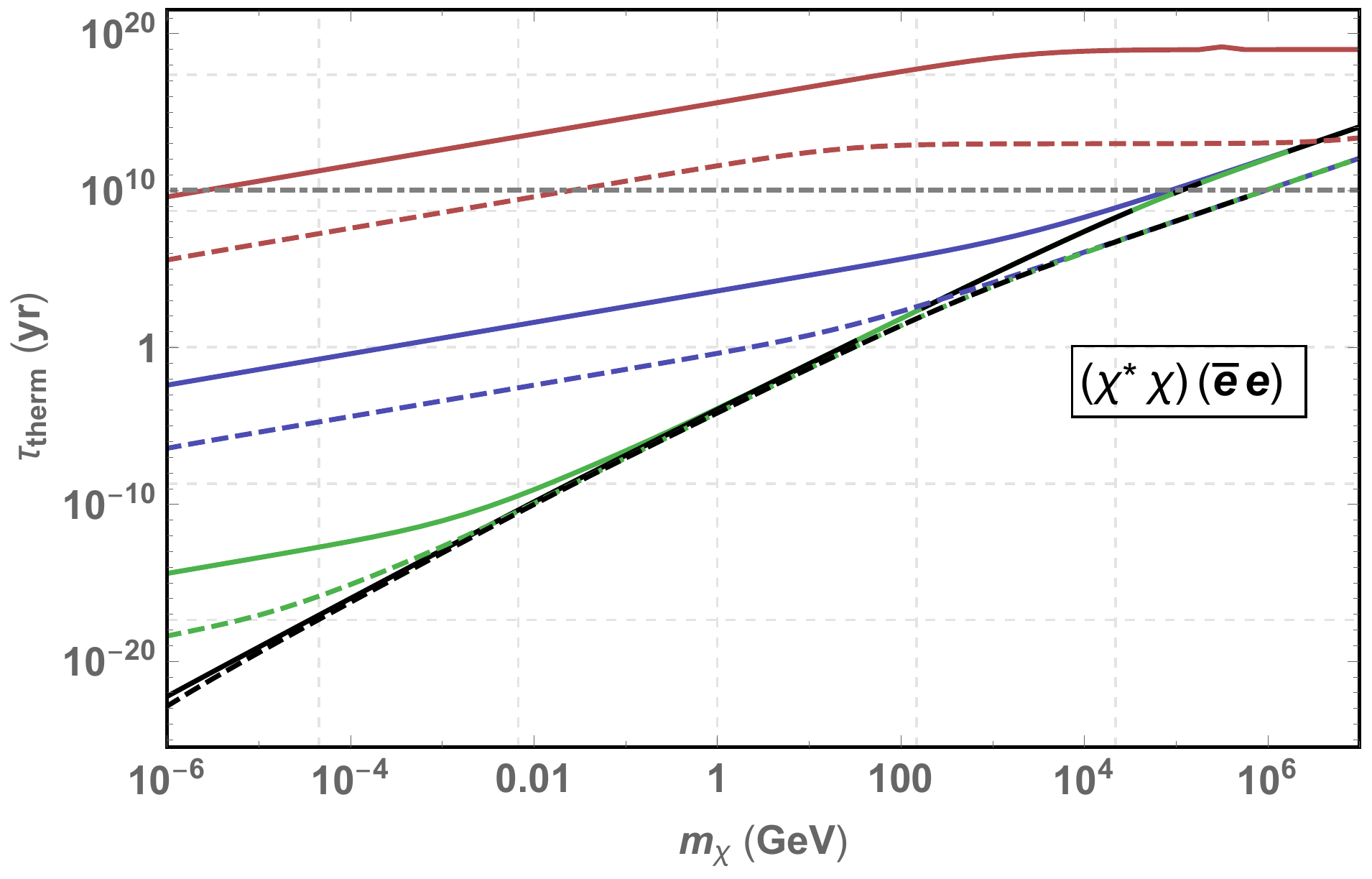}
    	\caption{Same as Figure~\ref{fig:tauthermnucleons}, but for {\bf spin-0 DM} and both {\bf neutron} (top panels) and {\bf electron} (bottom panel) targets.
    	For the operator $\mathcal{O}^\textnormal{S}_2$ the $\tau_{\rm therm}$ for electron scattering is the same as that of neutron scattering.
	}
	\label{fig:tauthermnspin0}
\end{figure*}
%%%%%%%

%%%%%%%%%
\begin{table*}
	\centering
	\begin{tabular}{llll}
		\hline
		Operator 
		& $m_{\phi} \gtrsim \sqrt{m_{\chi}\,T}$ 
		& $m_{\phi} \lesssim \sqrt{m_{\chi}\,T}$
		& $m_{\phi} \lesssim \sqrt{m_{\chi}\,T}$
		\\
		& & neutrons & electrons 
		\\
		\hline
		\hline
		\\
		$\mathcal{O}^F_{1}$
		&$\dfrac{35\pi^3}{12\,g_T^2 g_\chi^2\,T^2}\dfrac{m_\phi^4}{m_\chi\,m_T^2}$
		&$\dfrac{24\pi^3\,m_\chi}{g_T^2 g_\chi^2\,m_T^2}$
		&$\dfrac{10\pi^3\,m_\chi}{3g_T^2 g_\chi^2\,m_T^2}\ln\left(\dfrac{m_\chi^3 +3m_T^2m_\chi(2m_\chi/T+5)}{33m_T^2m_\chi+m_\chi^3}\right)$
		\\ \\
		\hline
		\\
		$\mathcal{O}^F_{2}$
		&$\dfrac{7\pi^3}{3\,g_T^2 g_\chi^2\,T^3}\dfrac{m_\phi^4}{m_T^2}$
		&$\dfrac{16\pi^3}{g_T^2 g_\chi^2\,T}\dfrac{m_\chi^2}{m_T^2}$
		&$\dfrac{20\pi^3\,m_\chi}{7g_T^2 g_\chi^2\,m_T^4T}\Bigg[7m_T^2(m_\chi-3T)+3T(m_\chi^2+2m_T^2)\times$
		\\
		& &
		&$\ln\left(\dfrac{3T(m_\chi^2+9m_T^2)}{7m_\chi m_T^2+3m_\chi^2 T+6m_T^2T}\right)\Bigg]$  
		\\
		\hline
		\\
		$\mathcal{O}^F_{3}$
		&$\dfrac{7\pi^3}{3\,g_T^2 g_\chi^2\,T^3}\dfrac{m_\phi^4}{m_\chi^2}$
		&$\dfrac{16\pi^3}{g_T^2 g_\chi^2\,T}$
		&$\dfrac{20\pi^3}{7g_T^2 g_\chi^2\,m_\chi T}\Bigg[7(m_\chi-3T)+9T \ln\left(\dfrac{30 T}{7m_\chi+9T}\right)\Bigg]$
		\\
		\hline
		\\
		$\mathcal{O}^F_{4}$
		&$\dfrac{55\pi^3}{36\,g_T^2 g_\chi^2\,T^4}\dfrac{m_\phi^4}{m_\chi}$
		&$\dfrac{140\pi^3 m_\chi}{3\,g_T^2 g_\chi^2\,T^2}$
		&$\dfrac{140\pi^3 m_\chi}{3\,g_T^2 g_\chi^2\,T^2}$
		\\
		\hline
		\hline
		\\
		$\mathcal{O}^S_{1}$
		&$\dfrac{35\pi^3}{3\,g_T^2 a_\chi^2\,T^2}\dfrac{m_\phi^4\,m_\chi}{m_T^2}$
		&$\dfrac{96\pi^3\,m_\chi^3}{g_T^2 a_\chi^2\,m_T^2}$
		&$\dfrac{40\pi^3\,m_\chi^3}{3g_T^2 a_\chi^2\,m_T^2}\ln\left(\dfrac{6(m_\chi/T) m_T^2+m_\chi^2 +14m_T^2}{m_\chi^2+32 m_T^2}\right)$
		\\
		\hline
		\\
		$\mathcal{O}^S_{2}$
		&$\dfrac{28\pi^3}{3\,g_T^2 a_\chi^2}\dfrac{m_\phi^4}{T^3}$
		&$\dfrac{64\pi^3\,m_\chi^2}{g_T^2 a_\chi^2\,T}$
		&$\dfrac{80m_\chi \pi^3}{7 g_T^2 a_\chi^2 \,T}\Bigg[7(m_\chi-3T)+6T \ln\left(\dfrac{27 T}{7m_\chi+6T}\right)\Bigg]$
		\\
		\hline
		\hline
	\end{tabular}   
		\caption{Limiting behaviours of thermalization time for the interaction structures in Table~\ref{tab:ampsq} for neutron and electron targets.
	         These expressions are good approximations to the numerically-obtained plots in Figs.~\ref{fig:tauthermnucleons}, \ref{fig:tauthermelectrons} and \ref{fig:tauthermnspin0}.
	         Here $T$ is the neutron star temperature.}
 \label{tab:fits}
\end{table*}
%%%%%%%%%%%%%

%%%%%%%%
\subsection{Interaction structures}
\label{subsec:interxnstructs}
%%%%%%%%

In Table~\ref{tab:ampsq} we provide the spin-averaged squared amplitudes for various contact operator interaction structures for both spin-$\half$ and spin-0 DM. 
These structures are chosen to span the possibilities of $|\mathcal{M}|^2$ dominated by various powers of the Mandelstam $t$, hence of the transfer momentum and energy.
This also results in a range of dominant dependences in the cross section on target spin and/or DM velocity at terrestrial direct detection experiments~\cite{Kumar_MatEl}.

We assume that these interactions arise from $t$-channel mediators of mass $\mmed$, thus the effective operator treatment is only valid for $\mmed$ exceeding the typical $q$ in the thermalization process.
For more general mediator masses we make the following substitutions in Table~\ref{tab:ampsq}.
For spin-$\half$ DM, we substitute $\Lambda^{-4} \ra g_\chi^2 g_T^2/(\mmed^2-t)^2$, where the $g_i$ are the mediator's couplings to $\chi$ and $T$.
We set these to a reference value $g_\chi = g_T = 10^{-5}$ throughout this paper.
For spin-0 DM, we substitute $\Lambda^{-2} \ra a_\chi^2 g_T^2/(\mmed^2-t)^2$, where $a_\chi$ is a trilinear coupling between $\chi$ and a spin-0 mediator.
We set these to a reference value $a_\chi$ = 0.1 MeV, $g_T = 10^{-5}$ throughout this paper.

We see that for our choice of operators the $|\mathcal{M}|^2$ are given by linear combinations of powers of $t$.
Thus the energy loss rate in Eq.~\eqref{eq:Elossrate} can be computed for separate powers of $t$, which becomes useful in the heavy mediator limit for non-relativistic targets, for which analytical expressions may be obtained.
We provide these expressions in Appendix~\ref{app:non-rel}.

%%%%%%%
\begin{figure*}
  \includegraphics[width=0.46\textwidth]{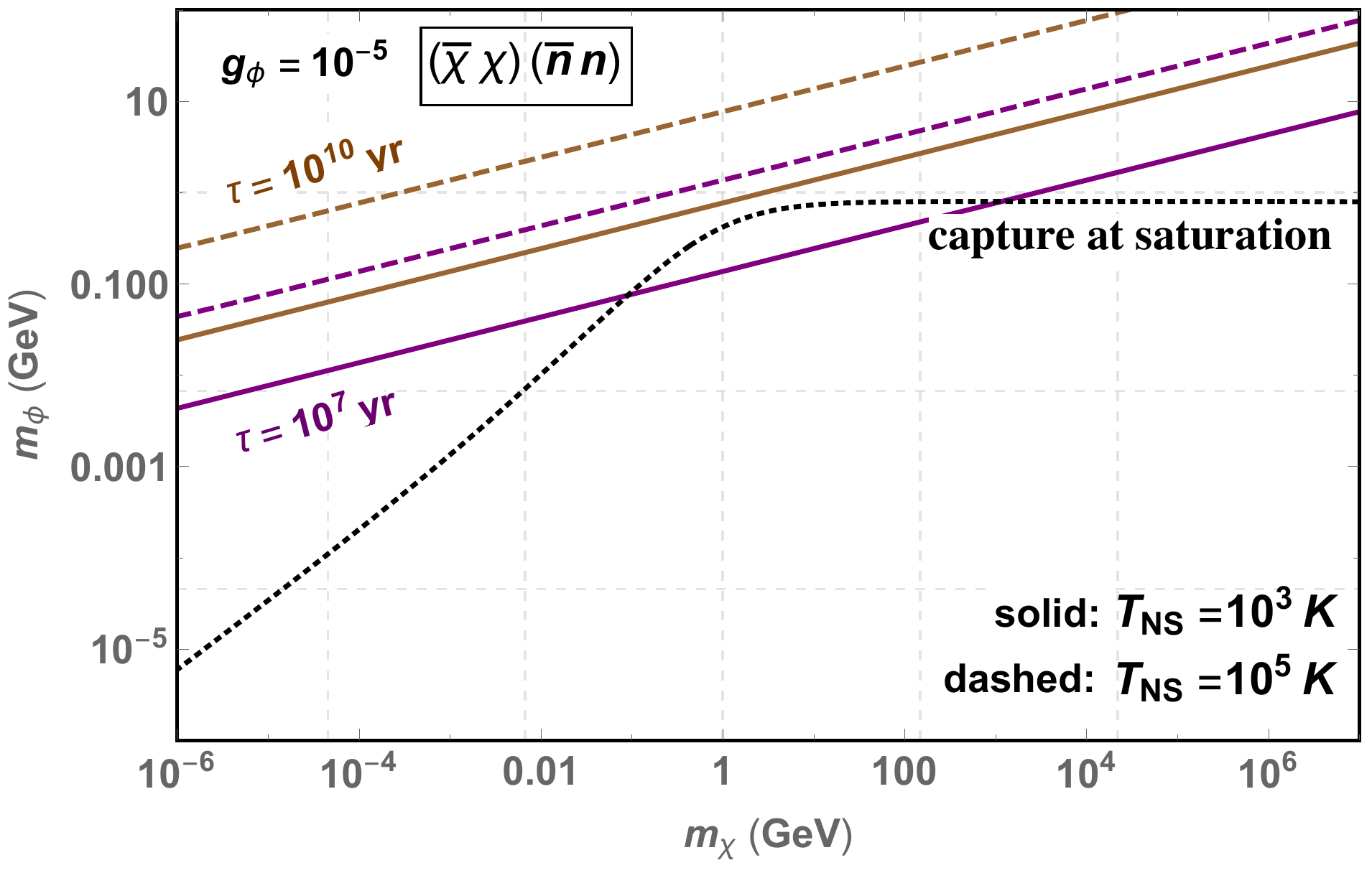}            
    \includegraphics[width=0.46\textwidth]{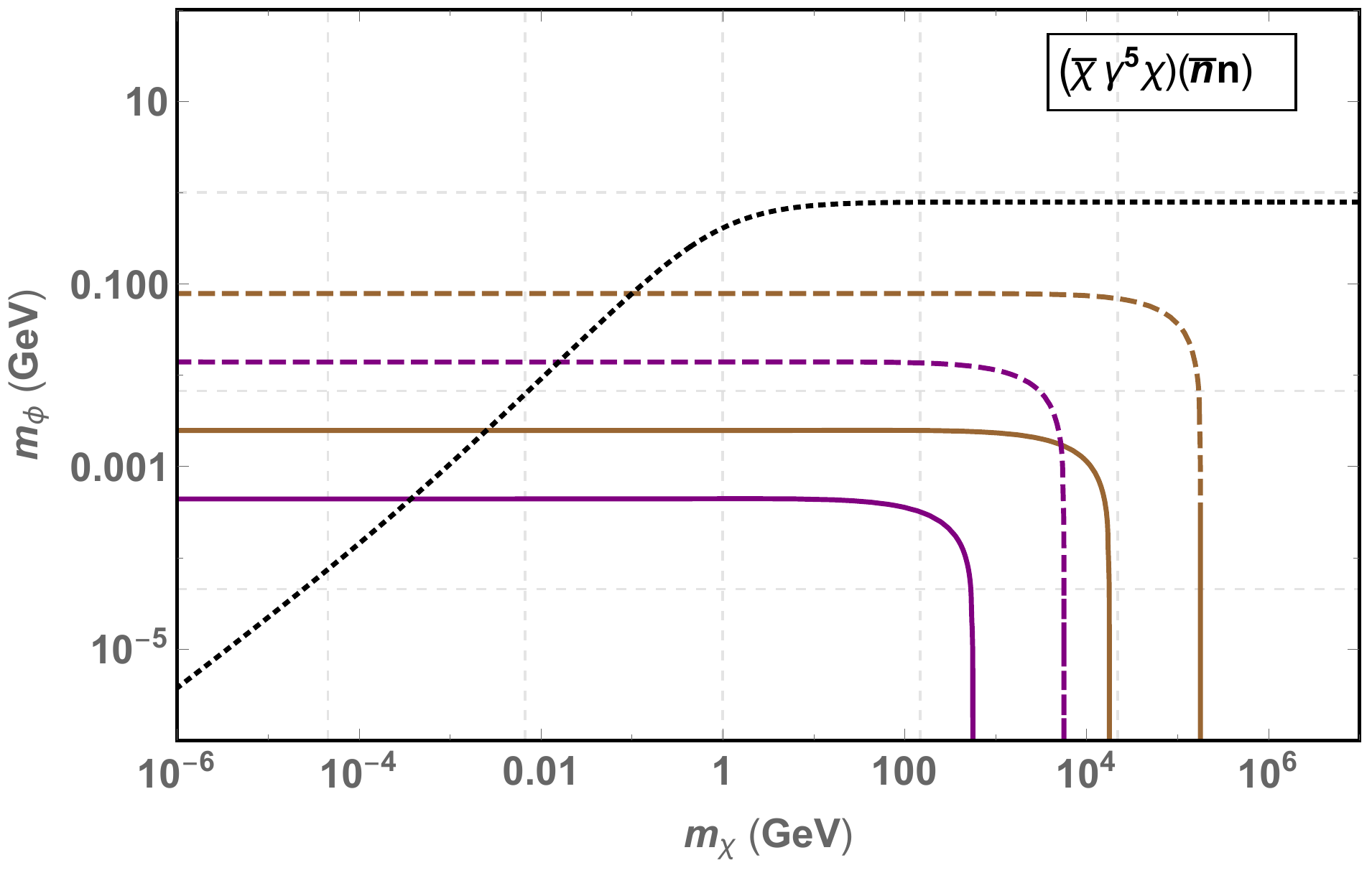} \\ 
    \includegraphics[width=0.46\textwidth]{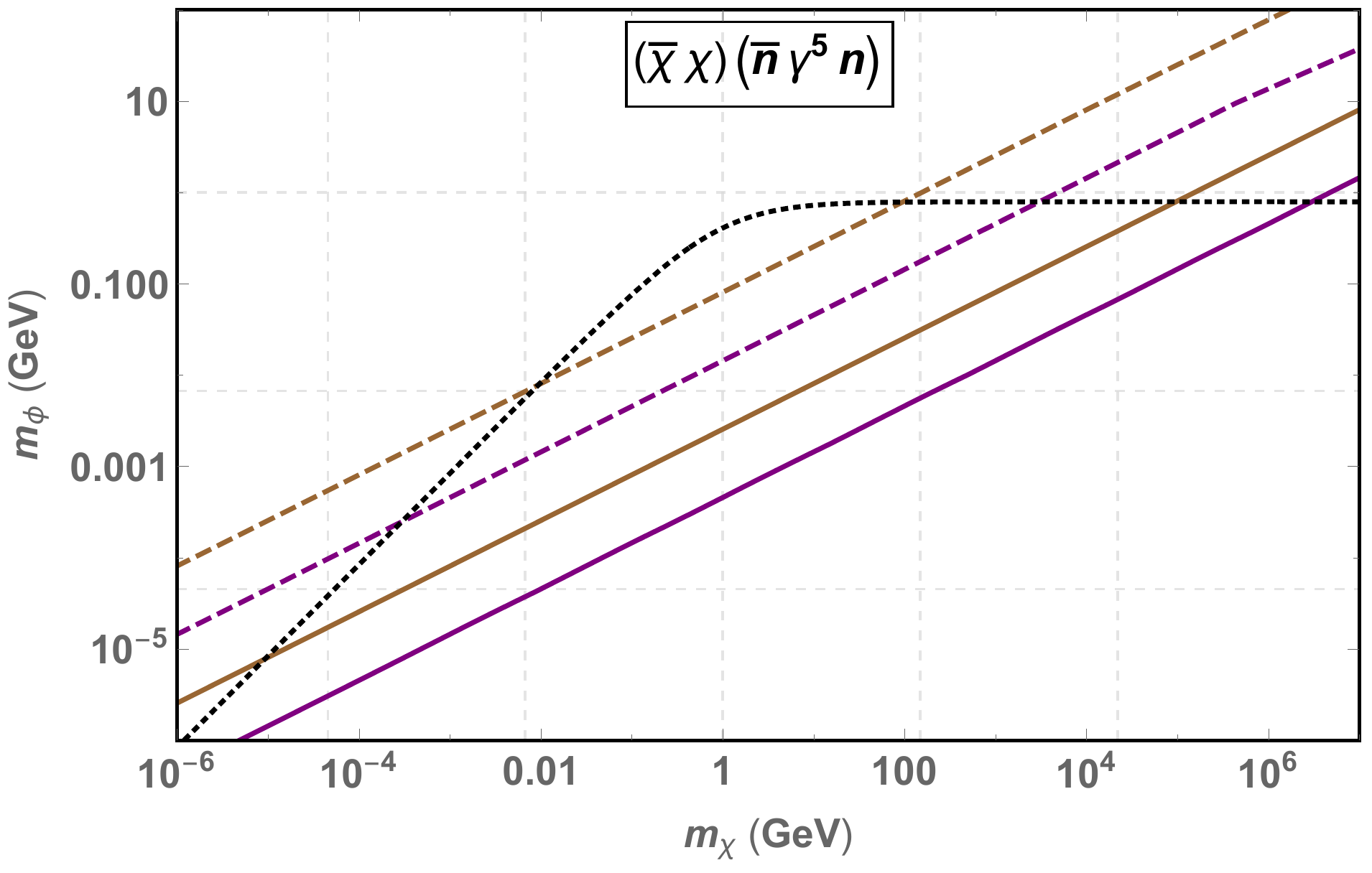}            
    \includegraphics[width=0.46\textwidth]{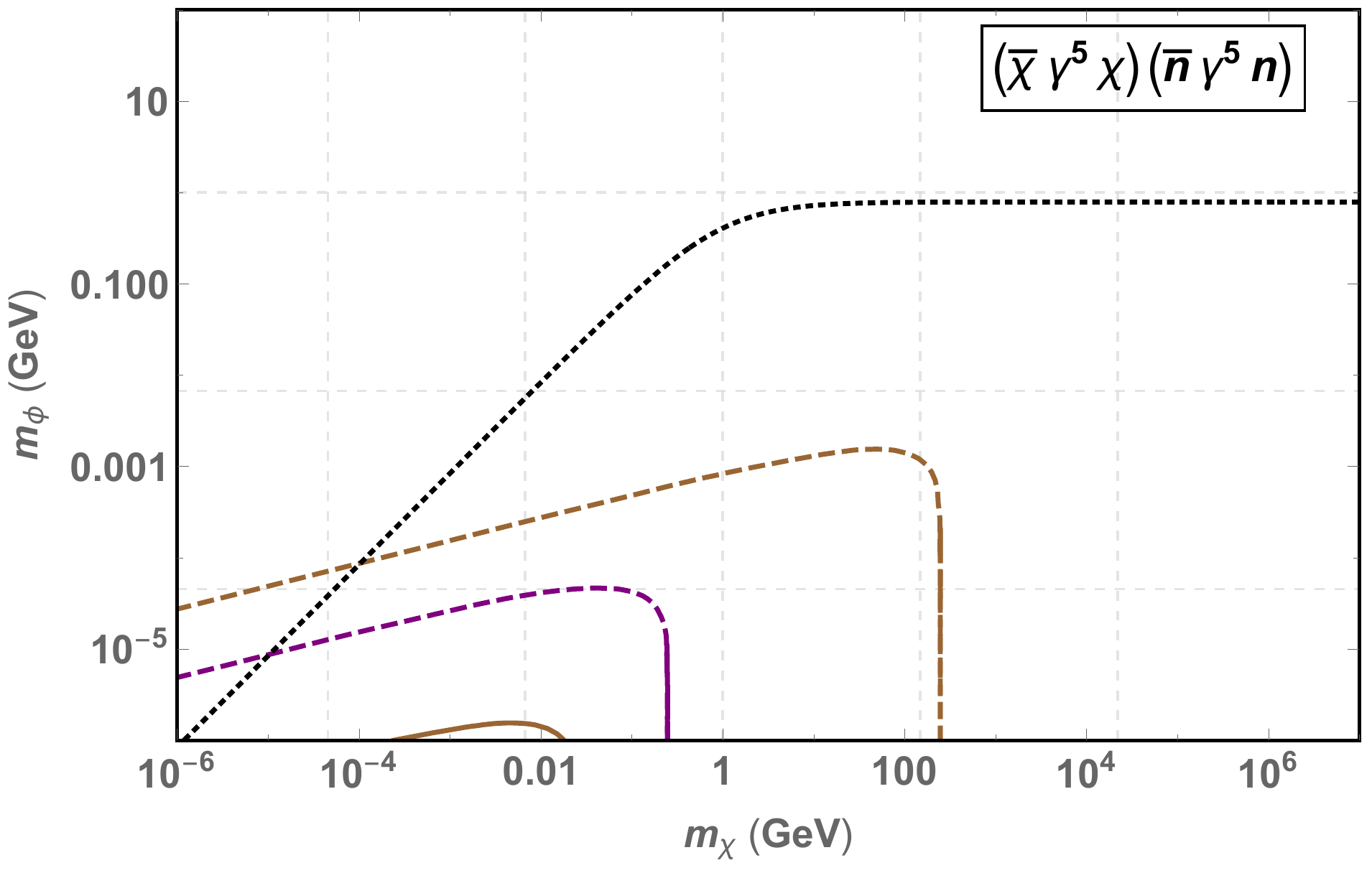} 
	\caption{Contours of constant thermalization time = 10$^7$ and 10$^{10}$ yr for NS temperatures 10$^3$ and $10^5$~K, for interactions between {\bf spin-$\half$ DM} and {\bf neutron} targets.
	Also shown is the curve corresponding to DM capture at saturation on a 1.5 $M_\odot$ mass, 12.6 km radius NS. 
	 The color code for all the panels is given by the first one.
    See text for further details.
	}
	\label{fig:limitnucleons}
\end{figure*}
%%%%%%%

%%%%%%%
\begin{figure*}
  \includegraphics[width=0.46\textwidth]{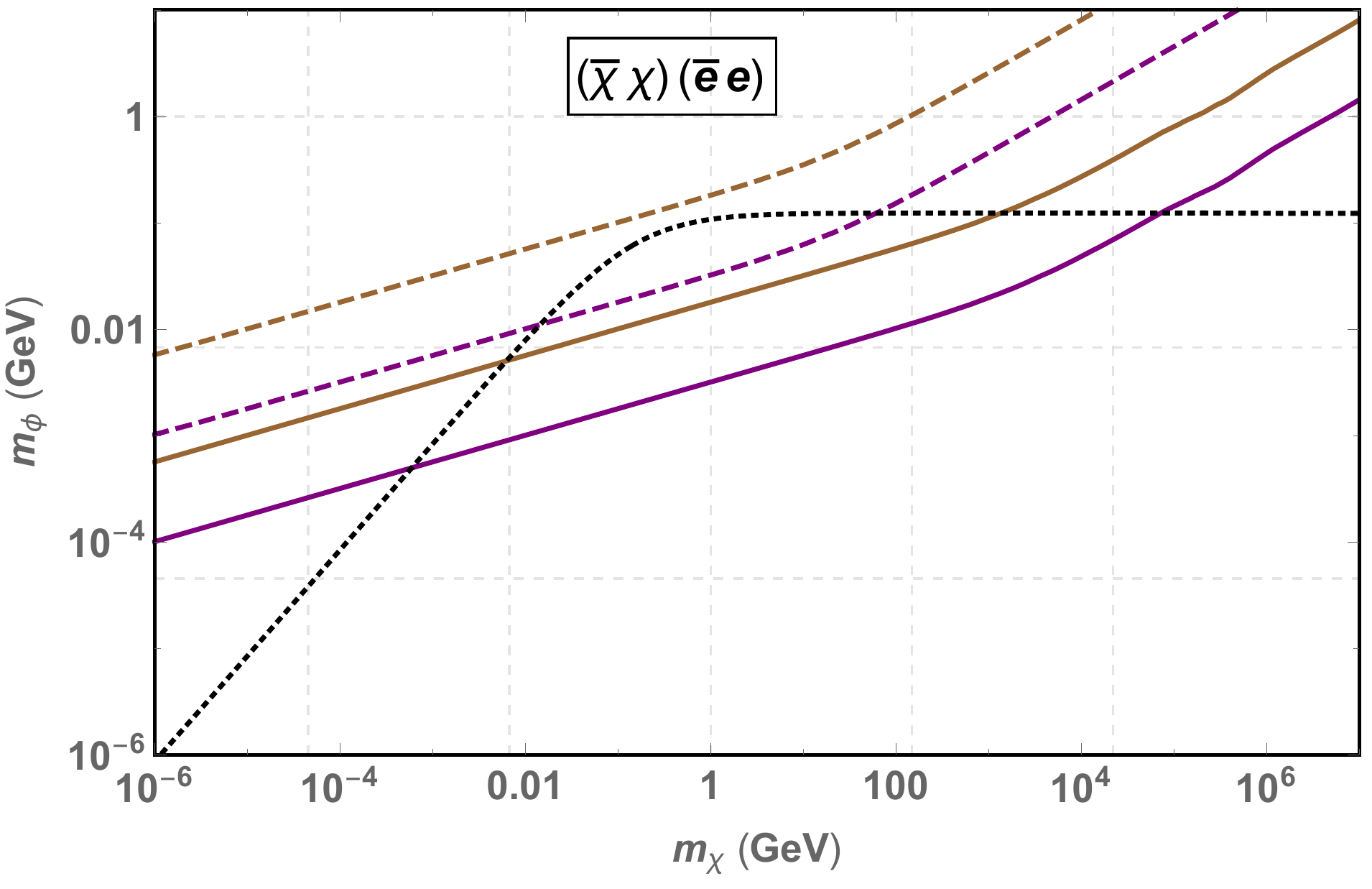}            
    \includegraphics[width=0.46\textwidth]{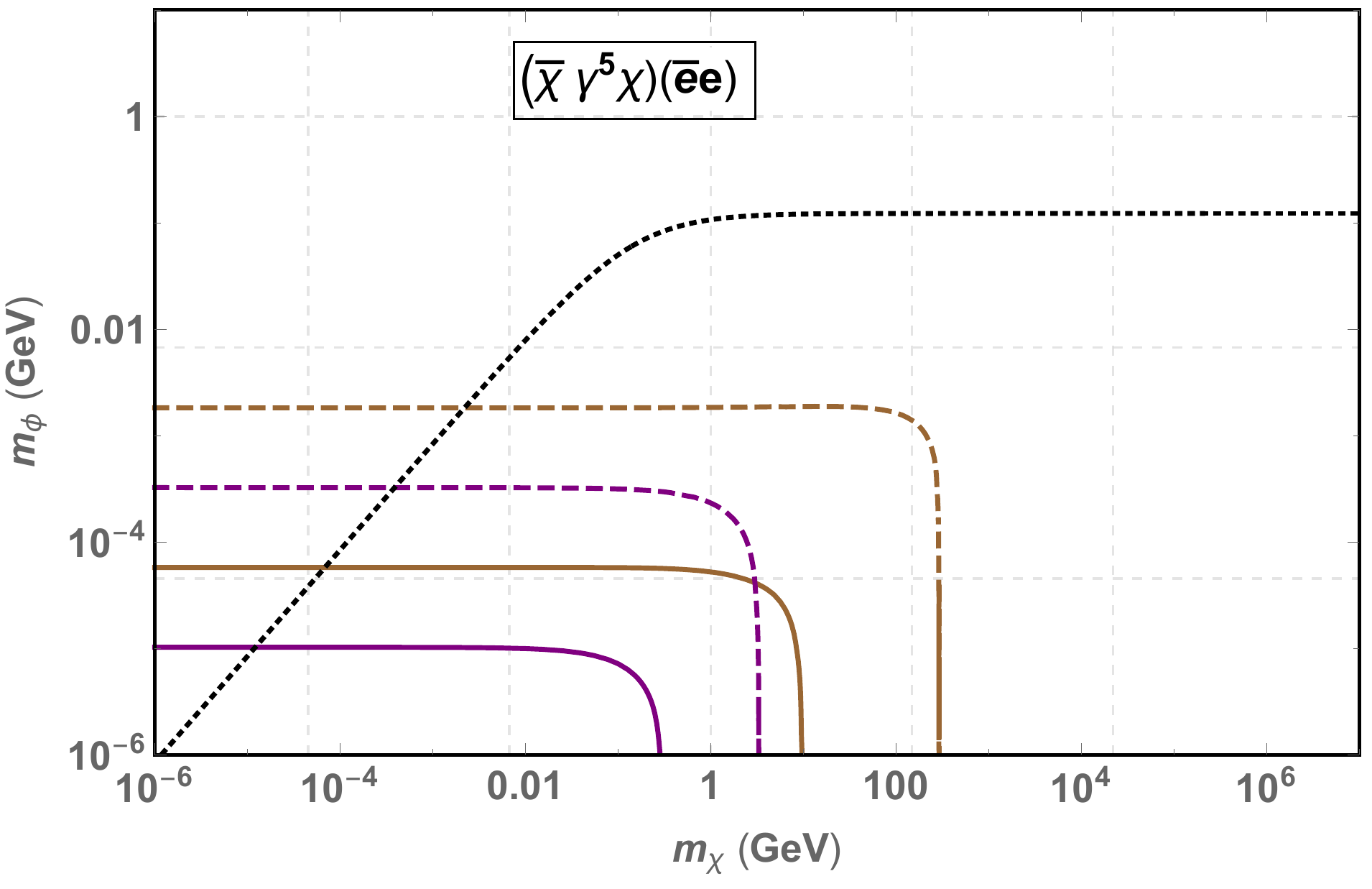} 
    	\caption{ Same as Figure~\ref{fig:limitnucleons}, but for {\bf spin-$\half$ DM} and {\bf electron} targets.
		}
	\label{fig:limitelectrons}
\end{figure*}
%%%%%%%

%%%%%%%
\begin{figure*}
  \includegraphics[width=0.46\textwidth]{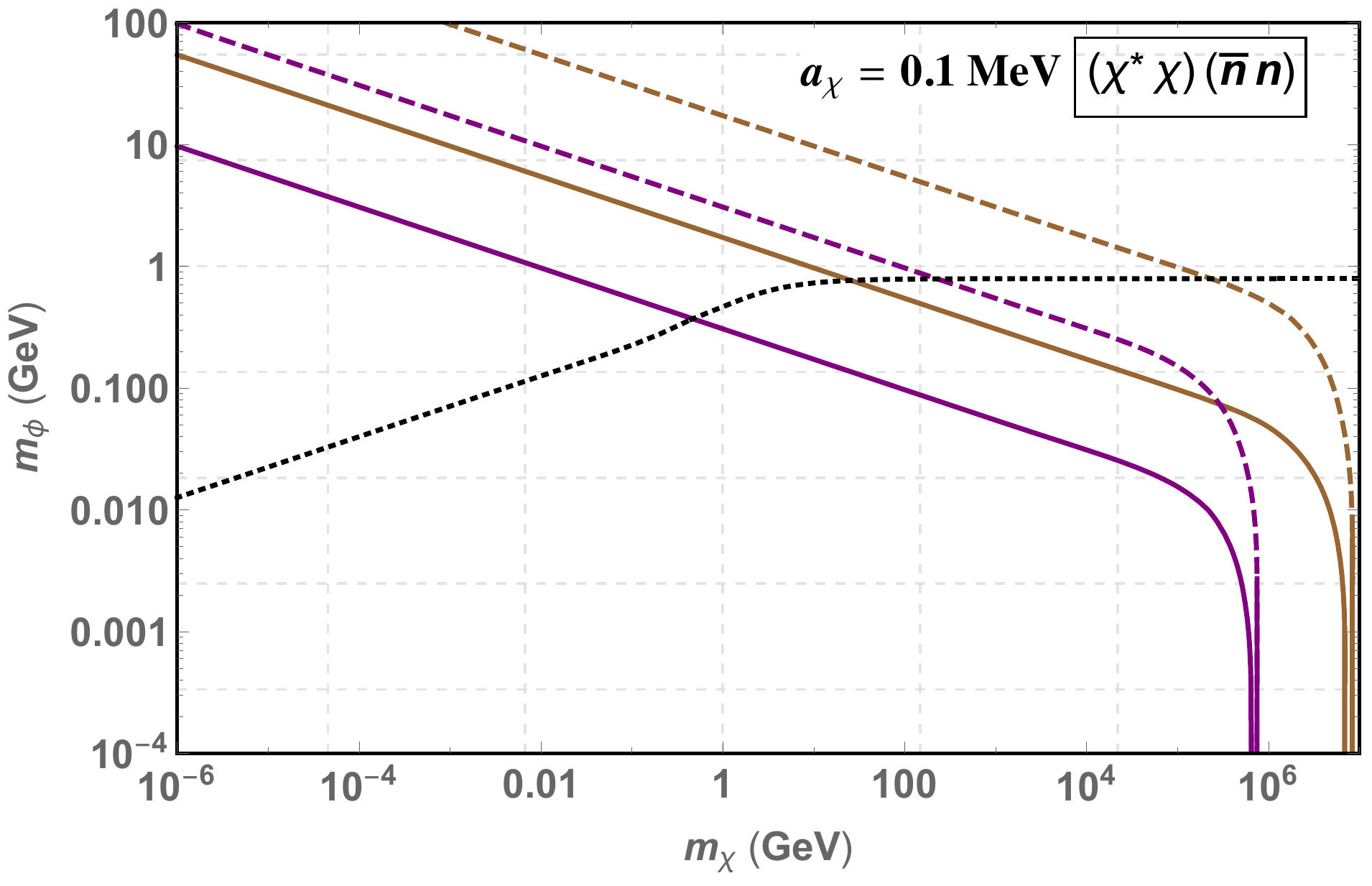}            
    \includegraphics[width=0.46\textwidth]{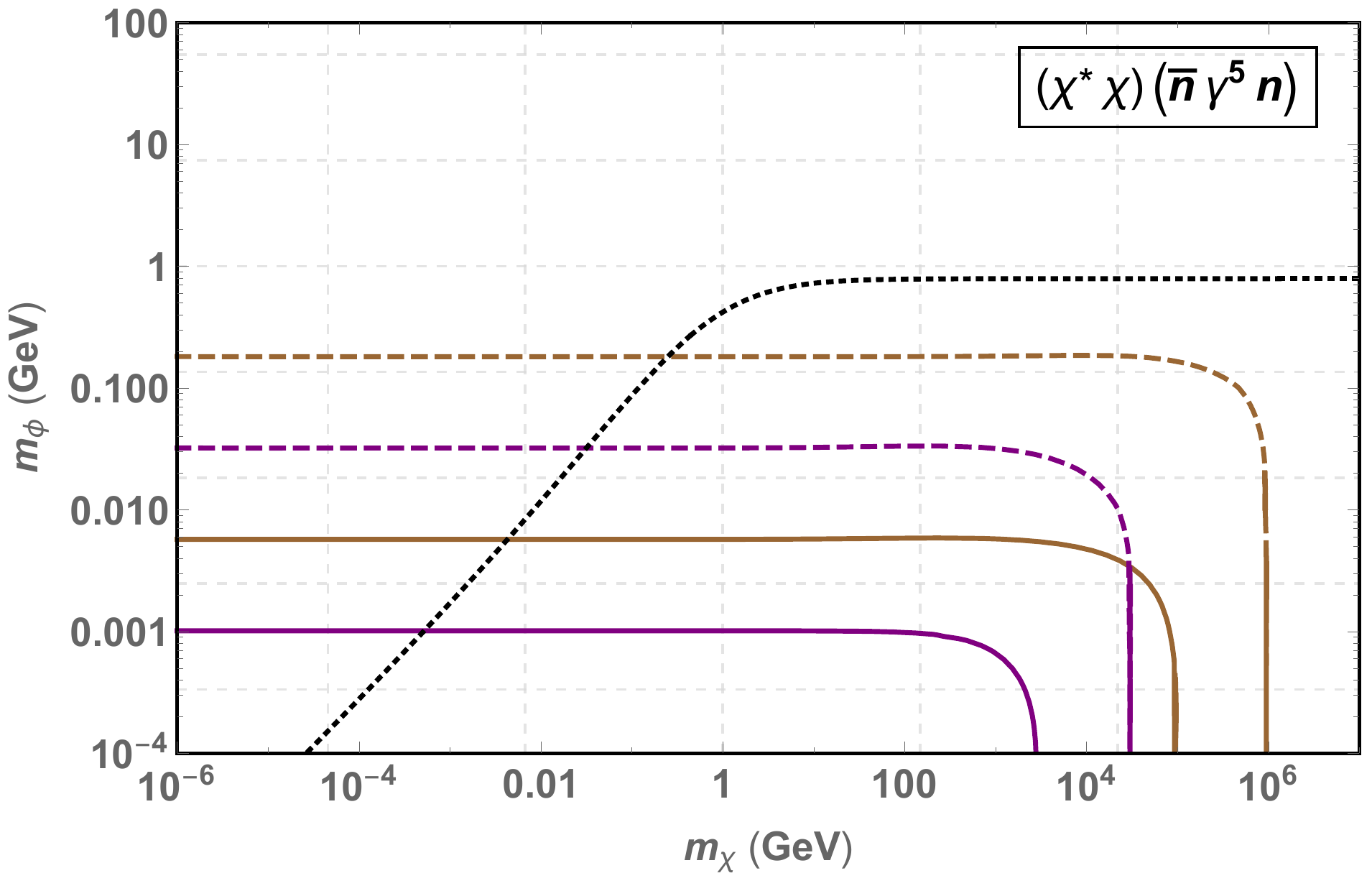} \\
    \includegraphics[width=0.46\textwidth]{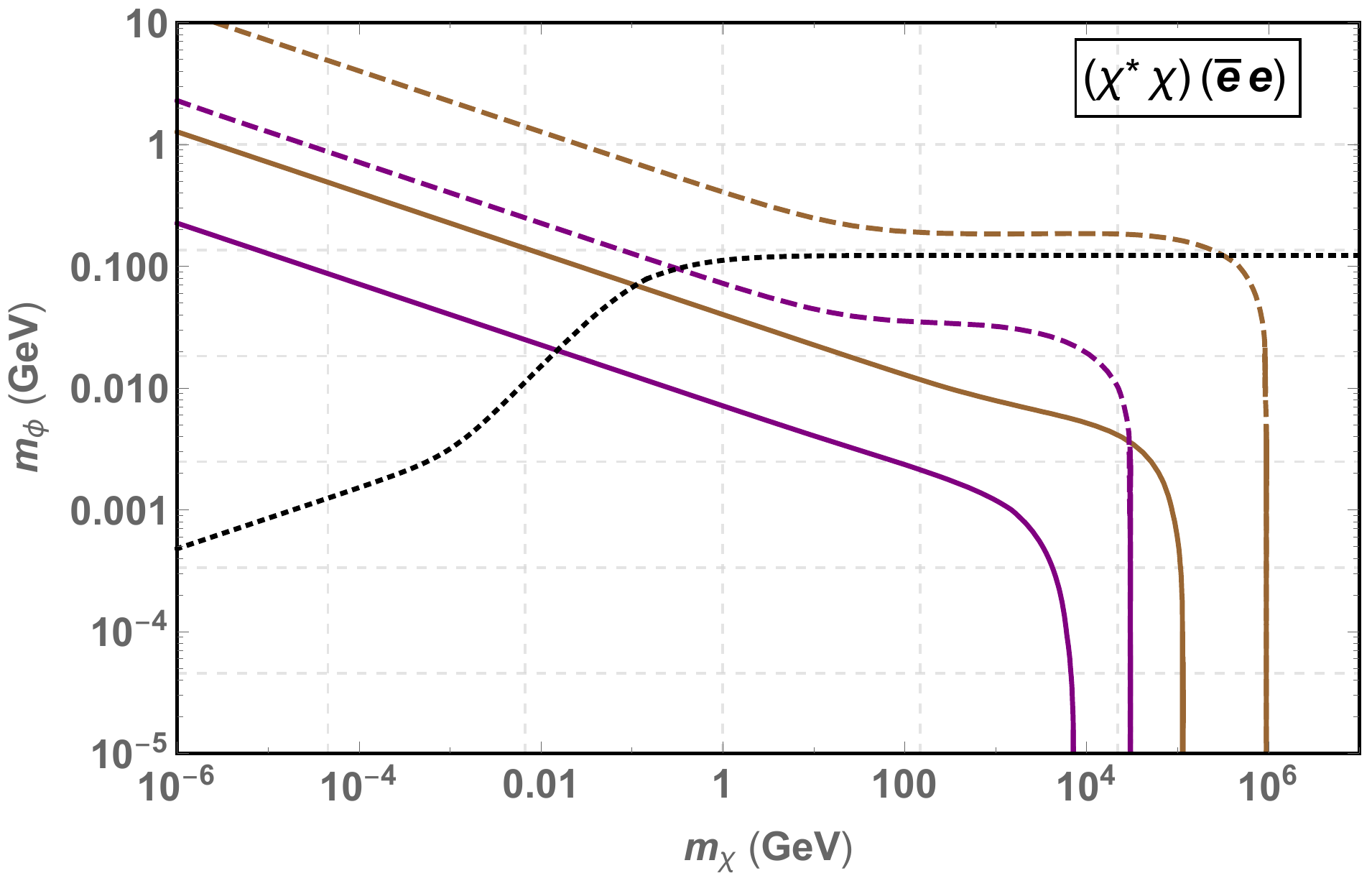} 
    	\caption{ Same as Figure~\ref{fig:limitelectrons}, but for {\bf spin-0 DM} and both {\bf neutron} (top panels) and {\bf electron} (bottom panel) targets.
	}
	\label{fig:limitspin0}
\end{figure*}
%%%%%%%

%%%%%%%%%
\section{Results}
\label{sec:results}
%%%%%%%%%

We now numerically compute the thermalization time using Eq.~\eqref{eq:ttime}, and display it as a function of $m_\chi$ for spin-$\half$ DM and neutron targets in Fig.~\ref{fig:tauthermnucleons}, 
for  spin-$\half$ DM and electron targets in Fig.~\ref{fig:tauthermelectrons}, and 
for spin-0 DM and both targets in Fig.~\ref{fig:tauthermnspin0}.
We show this for four choices of mediator masses: 
$\mmed =$~GeV, which serves as the ``heavy mediator" limit as it is well above the regime of $q$ relevant for DM-NS thermalization; 
$\mmed \to 0$, which serves as an asymptotic limit as we dial down the mediator mass; and 
$\mmed =$~keV and $\mmed =$~MeV to capture the intermediate behavior.
We also choose 2 NS temperatures down to which DM thermalizes: 
$10^5$~K, roughly the smallest observational upper bound placed on NS temperatures so far (see Table I of \cite{Yanagi_2020}), and 
$10^3$~K, roughly the temperature of DM kinetic and/or annihilation heating signals, to which upcoming IR telescopes are sensitive~\cite{Baryakhtar:2017dbj,Raj:2017wrv}.
We mark $\tau_{\rm NS} = 10^{10}$~yr as a benchmark for the oldest observable neutron stars. 
Hence parametric regions with $\tau_{\rm therm} > \tau_{\rm NS}$ are unlikely to have led to complete thermalization with the star.
In these regions DM could still have partially thermalized, i.e. not sunk entirely to the center of the star, and the ensuing DM annihilation (if allowed) could occur at non-trivial rates, and may or may not contribute to the observable heat signature.
We leave the investigation of these partial thermalization regimes to future work.
We do note that for electroweak-sized annihilation cross sections and DM masses, thermalization with just the km-thick crust of the NS is sufficient to result in the maximum stellar luminosity arising from DM annihilations~\cite{Acevedo:2019agu}.

We have omitted displaying the plots for electron targets for the operators $\mathcal{O}^\textnormal{F}_3$, $\mathcal{O}^\textnormal{F}_4$, and $\mathcal{O}^\textnormal{S}_2$ to avoid the following redundancy. 
As can be seen in Table~\ref{tab:ampsq}, their $|\mathcal{M}|^2$ is independent of the target mass, and thus the response functions for neutrons and targets are identical (Eqs.~\eqref{eq:response-nr-ex} and \eqref{eq:response-rel}), modulo the Heaviside theta function enforcing kinematic conditions. 
Since, as discussed, these conditions are satisfied in our parametric ranges, it effectively takes DM just as long to thermalize with the NS by scattering on electrons versus neutrons.
We choose the lower end of our $\mdm$ range to satisfy Gunn-Tremaine bounds on fermionic DM~\cite{Tremaine:1979we}; lower masses are allowed for spin-0 DM, but we do not display them here for brevity.

We notice serveral interesting features in these plots.
First, the slopes of the $\mmed =$~GeV and $\mmed \to 0$ curves are always distinctly different.
This is because the propagator term in $|\mathcal{M}|^2$ is dominated by $\mmed^{-4}$ in the first case and $q_{\rm ref}^{-4}$ in the second, where $q_{\rm ref} \sim \sqrt{m_\chi T_{\rm NS}}$ is the typical $q$ in the final few scatters that set the thermalization timescale.
This difference leads to different powers of $m_\chi$ feeding into the energy loss rate in Eq.~\eqref{eq:Elossrate} and thus ultimately into $\tau_{\rm therm}$ in Eq.~\eqref{eq:ttime}.
Second, we see that the keV and MeV mediator curves are parallel to the GeV mediator curve for small $\mdm$ -- here they behave as ``heavy mediators", with $\tau_{\rm therm} \propto \mmed^4$; these curves change slope and follow the massless mediator curve for large $\mdm$ -- here they are ``light mediators" with $\tau_{\rm therm} \propto q_{\rm ref}^4$. 
The turnaround point in $\mdm$ corresponds to $\mmed \simeq q_{\rm ref} = \sqrt{3 \mdm T_{\rm NS}}$.
We also see that while the $\mmed =$ GeV and $\mmed \to 0$ neutron curves are straight lines, the electron curves undergo changes in slope. 
This is because the $|\mathcal{M}|^2$ for neutron scattering is dominated by the term with the lowest power of $t$, giving simply power law results (see Appendix~\ref{app:non-rel}), whereas that for electron scattering is dominated by different powers in different $\mdm$ regimes due to relativistic kinematics.
Third, in small $\mdm$ regions where $|t| \ll m^2_{T, \chi}$, we find for the operators $\mathcal{O}^\textnormal{F}_1$, $\mathcal{O}^\textnormal{F}_2$, and $\mathcal{O}^\textnormal{S}_1$ that thermalization with electrons is a factor of $m^2_n/m^2_e$ longer than that with neutrons, as indeed expected from their $|\mathcal{M}|^2$, and Eqs.~\eqref{eq:response-nr-ex} \& \eqref{eq:response-rel}.
Fourth, comparing among the operators, we find that thermalization times are orders of magnitude longer for interactions whose $|\mathcal{M}|^2$ are dominated by more powers of $t$.
This is as expected, since the typical $|t|\simeq q^2 \simeq 3 \mdm T_{\rm NS}$ is much smaller than $m^2_{\chi,T}$, giving more suppressed $|\mathcal{M}|^2$ if dominated by higher powers of $t$.
Fifth, the heavy mediator limit curves for $\mathcal{O}^\textnormal{F}_2$ and $\mathcal{O}^\textnormal{S}_2$ are nearly flat versus $\mdm$.
This is because their $|\mathcal{M}|^2$ is dominated by $\mdm^2$ in the numerator, which cancels with the $\mdm^{-2}$ in the response function expression in Eqs.~\eqref{eq:response-nr-ex} \& \eqref{eq:response-rel}; the deviation from flatness for the $\mathcal{O}^\textnormal{F}_2$ electron curves  is due to competing effects from the $t^2$ term in the $|\mathcal{M}|^2$.
Finally, we see that colder NSs correspond to longer thermalization times, as expected.
The scaling of $\tau_{\rm therm}$  with temperature for $\mathcal{O}^\textnormal{F}_1$ is seen to agree with the analytical expression derived in Appendix~\ref{app:non-rel} for the heavy mediator limit and neutron targets.
In Table \ref{tab:fits} we provide the limiting behavior of $\tau_{\rm therm}$ in the heavy and light mediator limits for both neutron and electron targets; these expressions provide close analytical fits to the plots just examined.

The key features discussed above are reflected in our results identifying regions where DM-NS thermalization becomes important for observational prospects.
We turn to this next.

Our final results are displayed in Figs.~\ref{fig:limitnucleons}, \ref{fig:limitelectrons}, and \ref{fig:limitspin0}, corresponding respectively to thermalizing interactions between
spin-$\half$ DM \& neutrons, spin-$\half$ DM \& electrons, and spin-0 DM \& neutrons and electrons.
The ordering of the panels follows that of Figs.~\ref{fig:tauthermnucleons}, \ref{fig:tauthermelectrons}, and \ref{fig:tauthermnspin0}.
We show in the $\mmed$ vs $\mdm$ plane contours of $\tau_{\rm therm} = 10^7$~yr, 
corresponding to the age of a NS by which it is expected to have cooled down to $T_{\rm NS} \sim 10^3$~K~\cite{Yakovlev:2004iq,Page:2004fy},
and of  $\tau_{\rm therm} = 10^{10}$~yr, corresponding to the typical age of the oldest NSs.
As before, NS temperatures of 10$^3$~K and 10$^5$~K are chosen as our benchmark.

We indicate with a black dotted curve the mediator mass below which we expect all incident DM to be captured by our benchmark NS in Eq.~\eqref{eq:benchmarkNS}; for $\mmed$ larger than this ``saturation" limit, we expect a fraction of incident DM to capture.
For neutron targets this curve is obtained by equating the capture cross section computed in, e.g. Ref.~\cite{Baryakhtar:2017dbj}, to the interaction-dependent cross section $\sigma = \int d\cos\theta (32\pi s)^{-1} |\mathcal{M}|^2$, where we take $s \ra (m_T + \mdm(1+v^2_{\rm esc})/2)^2$ and $t \ra - 2 \mu^2_{\chi T} v^2_{\rm esc}$, with $\mu_{\chi T}$ the $\chi$-$T$ reduced mass.
For electron targets we obtain an approximate capture cross section by following the treatment of neutrons but replacing $m_n$ with $\mu_e$ and accounting for the reduced number of targets: electrons are 0.065 times as numerous as neutrons in our benchmark star.
Similarly we estimate the interaction-dependent cross section as above but with the replacement $m_n \ra \mu_e$. 
A more detailed estimate of capture via electrons obtained by following the treatment of Refs.~\cite{Joglekar:2019vzy,Joglekar:2020liw} results in a rate which differs from the above naive estimates by at most an order of magnitude.

The significance of the capture curve is that it depicts the regions where DM-NS thermalization becomes interesting from an observational viewpoint.
To illustrate with an example, suppose that a nearby, isolated neutron star were discovered, and its age determined to be $\sim 10^{10}$ yr. 
To place constraints on DM interactions based on the NS' thermal luminosity, we must first ask whether annihilations in the core are efficient, for which we check if the thermalization process has completed.
For $\mathcal{O}^\textnormal{F}_1$ and neutron targets, the $\tau_{\rm therm} = 10^{10}$ yr contour lies entirely outside the capture region, implying that thermalization is guaranteed in the candidate star. 
For all other operators and for electron targets, there are ranges of $\mdm$ -- prominently at low mass -- where the $\tau_{\rm therm} = 10^{10}$ yr contour lies inside the capture region. 
This implies that we cannot expect thermalization to have completed in the space between this contour and the dotted curve.

We see that the $\tau_{\rm therm}$ contours indeed agree with Figs.~\ref{fig:tauthermnucleons}--\ref{fig:tauthermnspin0} at the relevant mediator masses; the vertical fall-off in some of these curves correspond to the rapid turn-around and following-along of $\mmed \to 0$ curves in those plots.
Our results as a whole (Figs.~\ref{fig:tauthermnucleons}$-$\ref{fig:limitspin0}) give a representative picture of thermalization and its relationship with capture; similar results can be obtained by varying the benchmark couplings we had chosen in Sec.~\ref{subsec:interxnstructs}, but we do not expect the qualtitative conclusions to be different.
Parts of the parameter space in Figs.~\ref{fig:limitnucleons}$-$\ref{fig:limitspin0} are constrained by direct searches for DM scattering, indirect searches for galactic DM annihilation, the relic abundance of DM if set by thermal freeze-out, and collider searches.
These constraints, found in e.g. Refs~\cite{Raj:2017wrv,Bell:2019pyc,Joglekar:2019vzy,Joglekar:2020liw}, are not displayed here as the focus of our work is to explore regions of thermalization vis-a-vis capture in NSs.

%%%%%%%%%
\section{Discussion}
\label{sec:concs}

We have shown that, in several DM scenarios, thermalization in NSs post-capture is an important process that impacts observation prospects of candidate stars. 
 State-of-the-art infrared telescopes such as  the James Webb Space Telescope (JWST), Thirty Meter Telescope (TMT), and Extremely Large Telescope (ELT) are scheduled to be operational in the coming decade.
These are sensitive to extremely faint objects at 10 pc distances with blackbody surface temperatures down to $1000$ K~\cite{Raj:2017wrv}. 
In the so-called minimal cooling paradigm~\cite{Page:2004fy,Yakovlev:2004iq},
neutron stars (in the absence of exotic phases in the core) can ideally cool to temperatures well below $1000$ K within Gyr timescales. 
 On the other hand, DM can capture in NSs via scattering on its constituents, and the instantaneous transfer of kinetic energy can heat a typical NS to 1750~K in the solar neighbhorhood~\cite{Baryakhtar:2017dbj}. 
Additionally, if DM annihilations occur efficiently in the NS core, the star can heat up to a maximal temperature of $2500$ K, which could reduce telescope integration times by $\Oc(10)$ factors~\cite{Baryakhtar:2017dbj}.
But for successful annihilation-heating to happen, DM must first thermalize with the stellar core medium via repeated scattering within Gyr timescales, i.e. the typical ages of the oldest NSs.

In this work, we have computed the DM thermalization time for DM-SM interaction Lorentz structures listed in Table~\ref{tab:ampsq}, for non-relativistic (neutrons)  and relatvisitic (electrons) target particles, in both the effective ``heavy" mediator and ``light" mediator limits. 
We have also provided analytical functions for the thermalization time that reproduce well the asymptotic behavior of numerical results. 
We find that, except for the operator $\mathcal{O}^F_1$ with neutron targets, in all other scenarios there are regions where despite efficient capture of DM it does not thermalize with the NS within Gyr timescales. 
Thus DM will not maximally heat the NS via annihilations; for such regions, kinetic heating of NSs may be the only signature.
This effect appears at first glance to be a contradiction, since one would naively expect capture to be efficient in regions where thermalization is quick, and vice versa. 
However the dynamics of the two processes are governed by distinctly different scales.
Capture is determined chiefly by the escape energy, determined by macroscopic parameters of the star such as its mass and radius;  Pauli-blocking comes into play, in a simple manner, only for $\mdm$ below the Fermi momentum.
On the other hand, the thermalization timescale is set by the final kinetic energy of DM particles, i.e. the temperature of the NS core $T_{\rm NS}$. 
As discussed before, $q \ll \mu$ during the longest intervals of thermalization. 
In this regime, only those target particles close to their respective Fermi surfaces allow for DM to continuously lose energy and thermalize.

If DM thermalizes with the NS, the next question to address would be whether the DM capture and annihilation rates equilibrate. 
Once DM is thermalized, their distribution is confined to the thermal radius, given by~\cite{Bertoni:2013bsa}
%%%
\beq
r_{\rm th}=\left(\frac{3 \,T_{\rm NS} R_{\rm NS}^3}{ G M_{\rm NS} \mdm} \right)^{1/2} = 0.1\, {\rm m} \left(\frac{\rm GeV}{\mdm} \frac{T_{\rm NS}}{10^3\,{\rm K}}\right)^{1/2}~.
\eeq
%%%

Among the operators we have considered, $s$-wave annihilation to SM fermions dominates for $\mathcal{O}^{\rm F}_2$, $\mathcal{O}^{\rm F}_4$, $\mathcal{O}^{\rm S}_1$ and $\mathcal{O}^{\rm S}_2$, whereas the $p$-wave dominates for $\mathcal{O}^{\rm F}_1$ and $\mathcal{O}^{\rm F}_3$.
For sufficiently light mediators, DM annihilation to a pair of mediators will also be allowed. Below the electron kinematic threshold $\mdm \simeq 0.5$~MeV, only DM annihilations into $\gamma \gamma$ could possibly heat the NS; annihilations into neutrinos will deposit heat only for $m_\chi> 100$ MeV~\cite{Reddy:1997yr}.  
In order to estimate the typical annihilation cross section required for equilibration we parameterize $\langle \sigma v \rangle_{\rm ann} = a + b\,v^2$ and demand that $\tau_{\rm NS}/ \tau_{\rm eq} \lsim 5$, with the equilibration time $\tau_{\rm eq} = (V_{\rm th}/C \langle \sigma v \rangle_{\rm ann})^{1/2}$. 
Here $C$ is the DM capture rate and $V_{\rm th} = 4/3\, \pi r^3_{\rm th}$ is the thermal volume.
This results in the condition
%%%%
\beq
a> 7.5\times 10^{-54}\, {\rm cm^3/s} \left(\frac{\rm Gyr}{\tau_{\rm NS}}\right)^2 \left(\frac{C_{\rm sat}}{C}\right) \left(\frac{\rm GeV}{\mdm} \frac{T_{\rm NS}}{10^3\,{\rm K}}\right)^{3/2} \nonumber
\eeq
%%%%
when the annihilation is $s$-wave-dominated, and
%%%
\beq
b> 2.9\times 10^{-44}\, {\rm cm^3/s} \left(\frac{\rm Gyr}{\tau_{\rm NS}}\right)^2 \left(\frac{C_{\rm sat}}{C}\right) \left(\frac{\rm GeV}{\mdm} \frac{T_{\rm NS}}{10^3\,{\rm K}}\right)^{1/2}
\eeq
%%%%
when it is $p$-wave-dominated, and where $C_{\rm sat}$ is
the saturation capture rate.

 Current available data allow for some exotic phases such as hyperons and quark matter to perist in the NS core~\cite{Drischler:2016cpy,Alford:2017qgh,Alford:2020pld}. 
 As shown in Ref.~\cite{Bertoni:2013bsa}, DM does not thermalize with a NS core in a color-flavor-locked phase for vector operators. 
 Similar results hold for scalar operators.
 We note that in such a scenario only light DM, whose thermal radius extends beyond the core, can thermalize.
We also remark that when the core is dominated by exotic phases DM capture is likely effected by the NS crust~\cite{Acevedo:2019agu}, in which case the subsequent thermalization with the crust material, which is not Fermi-degenerate in most regions, can be treated in a more simple manner than done in our work.

%%%%%%%%%

%%%%%
\section*{Acknowledgments}
We are grateful to
Joe Bramante,
Thomas Hambye,
Ann Nelson,
Sanjay Reddy,
and
Peter Tinyakov
for beneficial conversations.
RG and AG are supported by the ULB-ARC grant ``Probing DM with Neutrinos'' and the Excellence of Science grant (EOS) convention~30820817.
The work of N. R. is supported by the Natural Sciences and Engineering Research Council of Canada. 
TRIUMF receives federal funding via a contribution agreement with the National Research Council Canada.

%%%%%%%%%

\appendix

%%%%
\section{Response function for relativistic (electron) targets}
\label{app:rel}
%%%%

For ultra-relativistic targets such as electrons we have $E_p = |p|$ and $E_{p'} = |p + q|$, so that the response function in Eq.~\eqref{eq:Sred} becomes
%%%
\bea
S(q_0,q) = && \frac{|\mathcal{M}|^2}{16\,m^2_\chi} \int \frac{\dd^3 p}{(2 \pi)^3 } (2 \pi) \delta\left(q_0 + |p| - |p +q| \right) \times \nonumber \\
&& \frac{1}{|p| |p +q| } f(|p|) \left(1 -f(|p +q|)  \right).
\eea
%%%
The delta function may be expressed in terms of the scattering angle $\theta_{pq}$:
\bea
\delta\left(q_0 + |p| - |p +q| \right) &=& - \frac{|p+q|}{p q} \delta\left(\cos\theta_{pq} - \cos\theta^*_{pq}\right) \times \nn \\
 && \Theta(p -p_-)~,\nonumber\\
\cos\theta^*_{pq}& =&  \frac{1}{2 p q} \left( q_0^2 + 2 p q_0 -q^2 \right)~,\nn\\
p_- &=& \half \left(q-q_0 \right).
\eea
Note that $f(|p +q|) \rightarrow f(p +q_0) $ upon integration over the delta function.
As the NS temperature we consider is $\mathcal{O}(0.1-10)$ eV, $q_0> T_{\rm NS}$ is a good approximation up to the last stages of thermalization. 
In the limit $\mu \gg T_{\rm NS}$ in the above integral, we have  
%%%
\bea
\int^{\infty}_{p_{-}} \dd p f(p) \left(1 -f(p +q_0)  \right) &=&  \frac{q_0}{1 -e^{-q_0/T_{\rm NS}} }  \times\nonumber \\
&&\left(-1 + \Theta(-q_0 + q - 2 \mu) \right)\nonumber \\
&\ra& q_0 \Theta(2 \mu +   q_0 - q )~.
\eea
%%%
Using this we obtain the response function for relativistic targets written in Eq.~\eqref{eq:response-rel}.

%%%%
\section{Analytic expressions for energy loss rate for non-relativistic (neutron) targets}
\label{app:non-rel}
%%%%

In this appendix we provide analytic expressions for the thermalization time for neutron targets in the heavy mediator limit.
These are obtained by writing out the energy loss rate in Eq.~\eqref{eq:Elossrate} as
%%%%
\bea
\label{eq:etransfer}
\Phi & =& \frac{1}{2\,\pi^2}\int k^{\prime \,2} d k^\prime d \cos \theta_{k k^\prime} S(q_0,q)\nonumber\\
&& \times \left(\frac{k^2}{2 \,m_\chi}-\frac{k^{\prime\,2}}{2 \,m_\chi} \right)~,
\eea
%%%%
where $S(q_0,q)$ is given by Eq.~\eqref{eq:response-nr-ex}.

From Table~\ref{tab:ampsq}, for mediator masses $\gg q_{\rm ref} \sim \sqrt{3\mdm T_{\rm NS}}$, we can expand the squared amplitude as:
%%%%
\bea
|\mathcal{M}|^2 = \sum_{n=0,1,2} \alpha_n t^n~,
\eea
%%%%
where the coefficients $\alpha_n$ can be read off of Table~\ref{tab:ampsq}.
We can then break down the response function into powers of $t = (q_0^2 - q^2)$:
%%%%
\begin{equation*}
    S_n(q_0,q)= \frac{\alpha_n} { 16 \pi m^2_\chi } \frac{q_0}{q} t^n~.
\end{equation*}
%%%%
This allows us to break down the energy loss rate as well (Eq.~\eqref{eq:etransfer}), and we get:
%%%
\bea
\nn \Phi_0 &=& \tilde{\alpha}_0 \frac{ \,k^6}{105\, m^2_\chi},\\
 \Phi_1 &=& \tilde{\alpha}_1 \frac{2 \,k^8}{63 m^2_\chi}\left(\frac{1}{3}-\frac{2 \,k^2}{55\, m^2_\chi} \right),\\
\nn \Phi_2 &=& \tilde{\alpha}_2 \frac{8 \,k^{10}}{495 m^2_\chi}\left(1 - \frac{2\, k^2}{13\,m^2_\chi} + \frac{k^4}{91\,m^4_\chi} \right)~,
\label{eq:Phibreakdown}
\eea
%%%
where $\tilde{\alpha}_n=\alpha_n/ (16 \pi^3 m^2_\chi) $. 

We note that using the expression for $\Phi_0$ above in Eq.~\eqref{eq:ttime}, and with $k_{\rm hot} = \mdm v_{\rm esc}$ and $k_{\rm cold} = \sqrt{3 \mdm T_{\rm NS}}$, we obtain 
%%%
\beq
\tau_{\rm therm} = \Lambda^4 \frac{105\pi^3}{4}\frac{\mdm}{m^2_n}\bigg(\frac{1}{k^4_{\rm hot}} - \frac{1}{k^4_{\rm cold}}\bigg)~,
\eeq
%%%
in agreement with Ref.~\cite{Bertoni:2013bsa}.
This also agrees with the first entry in Table~\ref{tab:fits}.

More generally, Eq.~\eqref{eq:Phibreakdown} implies that at leading order in $k_{\rm cold}$, we have $\tau_{\rm therm} \propto k^{-6}_{\rm cold} \propto T_{\rm NS}^{-3}$ for $\mathcal{O}^\textnormal{F}_2$, $\mathcal{O}^\textnormal{F}_3$ and $\mathcal{O}^\textnormal{S}_2$, and $\tau_{\rm therm} \propto k^{-8}_{\rm cold} \propto T_{\rm NS}^{-4}$ for $\mathcal{O}^\textnormal{F}_4$, as reflected in Figs.~\ref{fig:tauthermnucleons}$-$\ref{fig:tauthermnspin0} and Table~\ref{tab:fits}.

%%%%%
\section{Number of scatters to thermalize}
\label{app:Nscatt}
%%%%%%%%

It is instructive to estimate the number of scatters required for DM-NS thermalization.
Assuming the squared amplitude is independent of Mandelstam variables, from Eq.~\eqref{eq:rate}
the average energy lost in a collison for neutron targets is given by
%%%
\bea
\left\langle \Delta E \right\rangle^{\rm non-rel}  = \frac{\int_0^k d\Gamma(E_i) \,(E_i-E_f)}{\int_0^k d\Gamma(E_i)} = \frac{4}{7}E_i~,
\label{eq:delta_Enr}
\eea
%%%
where $E_i$ is DM energy before collision. 
For electron targets we similarly have
%%%
\beq
\left\langle \Delta E \right\rangle^{\rm rel}  \approx \frac{2}{3}E_i~.
\label{eq:delta_Erel}
\eeq
%%%

These fractional energy losses are huge, hence we can expect DM to thermalize after a small number of scatters.
This is obtained as the sum of a geometric series:
%%%
\bea
 \mathcal{N}_{\rm T} = \frac{\log (E_{\rm th}/E_0)}{\log \left(1- \alpha_{\rm T} \right)}~.
 \label{eq:number_scat}
\eea
%%%

Here the thermal energy $E_{\rm th} = 3/2 \,T_{\rm NS}$,
the initial DM energy $E_0 =  m_\chi v_{\rm esc}^2/2$,
and $\alpha_{\rm T} =$ 4/7 (2/3) for neutron (electron) targets. 
Thus for $T_{\rm NS} = 10^3-10^5$~K, $\mathcal{N}_{\rm T} = \mathcal{O}(10^1-10^2)$ over the whole range of DM masses considered, with a gradually increasing logarithmic dependence on the mass.
In particular, for $T_{\rm NS} = 10^3$~K, we find that $\mathcal{N}_{\rm n}$ spans $9-44$ and $\mathcal{N}_{\rm e}$ spans $7-34$.
For squared amplitudes that depend on $t^n$, $n \geq 1$, we find again that the number of scatters is $\mathcal{O}(10)$, with $\Oc(1)$ variation with respect to the above.

\bibliography{refs}

\end{document}

%% file: universalnewcommands.tex
\newcommand{\half}{\frac{1}{2}}

\newcommand{\lsim}{\lesssim}
\newcommand{\ra}{\rightarrow}

\def\Oc{\mathcal{O}}

\renewcommand{\tilde}{\widetilde} 
\newcommand{\beq}{\begin{equation}}
\newcommand{\eeq}{\end{equation}}
\newcommand{\bea}{\begin{eqnarray}}
\newcommand{\eea}{\end{eqnarray}}
\newcommand{\nn}{\nonumber}